\documentclass{article}
\usepackage[utf8]{inputenc}
\usepackage{natbib}
\usepackage{amsmath,amsfonts,amssymb,amsthm}
\usepackage{graphicx}
\usepackage{rotating}
\usepackage{caption}
\usepackage{subcaption}
\usepackage{authblk}
\usepackage{amsthm}
\usepackage{pgf,tikz}

\usetikzlibrary{positioning}

\newcounter{subassumptionR}[assuR]

\makeatletter
\renewcommand{\p@subassumptionR}{\theassuR}
\makeatother

\renewcommand\theassuR{R\arabic{assuR}}

\newcounter{subassumptionG}[assuG]

\makeatletter
\renewcommand{\p@subassumptionG}{\theassuG}
\makeatother

\renewcommand\theassuG{G\arabic{assuG}}

\newcounter{subassumption}[assu]

\makeatletter
\renewcommand{\p@subassumption}{\theassu}
\makeatother

\renewcommand\theassu{\arabic{assu}}

\newcounter{subdefinition}[deff]

\makeatletter
\renewcommand{\p@subdefinition}{\thedeff}
\makeatother

\renewcommand\thedeff{\arabic{deff}}

\newcommand{\ddml}{$D^2ML$ }
\newcommand{\ddmllong}{\textit{Dominant Drivers by Machine Learning}}

\title{Dominant Drivers of National Inflation\footnote{We are grateful for many comments from participants of the 19th International Workshop on Spatial Econometrics and Statistics, 27th International Conference on Computing in Economics and Finance, 2022 Panel Data Conference, seminar participants from Lund University and BI Norwegian Business School. We also appreciate helpful comments from Yigit Aydede, Vasilis Sarafids. The Authors acknowledge financial support from Italian Ministry MIUR under the PRIN project Hi-Di NET - Econometric Analysis of High Dimensional Models with Network Structures in Macroeconomics and Finance (grant 2017TA7TYC).}}

\author[1]{Jan Ditzen\thanks{jan.ditzen@unibz.it, www.jan.ditzen.net}}
\author[1]{Francesco Ravazzolo\thanks{francesco.ravazzolo@unibz.it, www.francescoravazzolo.com}}
\affil[1]{Free University of Bozen-Bolzano, Italy}

\date{\today}
\begin{document}
\maketitle

\begin{abstract}
For western economies a long-forgotten phenomenon is on the horizon: rising inflation rates. We propose a novel approach christened \ddml to identify drivers of national inflation. \ddml combines machine learning for model selection with time dependent data and graphical models to estimate the inverse of the covariance matrix, which is then used to identify dominant drivers. Using a dataset of 33 countries, we find that the US inflation rate and oil prices are dominant drivers of national inflation rates. For a more general framework, we carry out Monte Carlo simulations to show that our estimator correctly identifies dominant drivers.
\newline 
\end{abstract}

\section{Introduction}

The late 1990s marked the start of a period with low inflation rates across the world \citep{Rogoff2003}. The only exception were the years around the Great Financial crisis. The emergence of the COVID19 pandemic and related supply chain issues brought inflation back to the headlines. The Russian invasion of Ukraine and the resulting increases in energy prices further fuelled inflation, especially in western countries. While it is well understood that national inflation is driven by national factors \citep{Auer2019}, the effects of spillovers across countries are less researched. \cite{Auer2019} find that international input linkages synchronize inflation rates between countries. \cite{Bataa2013} present evidence that the Euro Area leads inflation in North America and that inflation rates are more synchronized since the 1980s. The effect of global factors or common factors on national inflation is well understood. \cite{Ciccarelli2010} establish that 2/3 of national inflation is due to global inflation. However global inflation is not a stand in for common shocks such as changes in commodity prices and they find that no country is leading global inflation. 

To fill this gap in the literature, we use a novel approach to identify dominant drivers influencing national inflation using the GVAR database \citep{MehdiRaissi2020}. Drivers can be other countries' inflation rate or macroeconomic variables of the same or other countries.\footnote{Dominant drivers are often labelled as dominant series or units. \cite{Brownlees2021} call them \textit{granular series} and \cite{KapetaniosPesaranReese2020} call them \textit{units with pervasive effects}. Throughout the paper we will refer to them as \textit{dominant drivers}.} We find that the inflation rates in the United States and oil prices changes have a dominant effect on national inflation rates in a set of 33 countries. Our results are robust to different estimation methods and specifications. An advantage of our approach is that it allows to identify dominant drivers even if the number of variables is larger than the number of observations over time.

Our approach relies on the inverse of the covariance matrix of the data and consists of two steps. The first step is based on \cite{Meinshausen2006} and \cite{Sulaimanov2016} and uses a graphical model to estimate the inverse of the covariance matrix the precision or concentration matrix. In a graphical model nodes (variables) are connected by edges (connections). A zero in the concentration matrix indicates independence between the variables, in a graphical model no connection. A non-zero in the concentration matrix implies dependence or a common edge between two nodes in the graphical model. Hence estimating the linkages between units using a graphical model is informative about the structure or sparseness of the concentration matrix. The entries of the concentration matrix can be represented by partial correlations and are therefore related to estimated regression coefficients \citep{Sulaimanov2016}. \cite{Meinshausen2006} and \cite{Sulaimanov2016} propose to use the least absolute shrinkage and selection operator (LASSO) estimator to estimate the graphical model and then use post LASSO OLS to estimate the elements of the concentration matrix. 
We further extend the approach by \cite{Meinshausen2006} and \cite{Sulaimanov2016} to time dependent data by combining it with either the rigorous LASSO \citep{Bickel2009a,Belloni2016,Chernozhukov2019,AhrensAitkenDitzenEtAl2020} or the adaptive LASSO estimator \citep{Zou2006,Medeiros2016a}. The advantage is that our approach can be applied to examples where the time dimension is smaller than the number of series or units.

The second step is the selection of the dominant drivers. We use the procedure in \cite{Brownlees2021}, henceforth BM, to select the dominant drivers from the estimated concentration matrix. In the BM procedure, the column norms of the concentration matrix are ordered by their size and the dominant drivers identified using a criterion similar to the eigenvalue ratio criterion  in \cite{AhnHorenstein2013}. We propose to use the heteroskedastic and autocorrelation robust rigorous LASSO and the adaptive LASSO to estimate the graphical model. Based on the linkages, post-LASSO OLS estimates the entries of the concentration matrix. Monte Carlo simulation results show that our proposed extension correctly identifies the dominant drivers. Since our approach relies on machine learning methods to identify the dominant drivers, we call it \ddmllong, \ddml.

The theoretical contribution of \ddml is threefold: First, it can be applied to data which has more variables than observations, a disadvantage of the method by \cite{Brownlees2021}. Secondly it is computationally more simple than the method in \cite{KapetaniosPesaranReese2020} and requires less assumptions. Finally, it can be applied to various types of data as shown in our Monte Carlo Simulations.

Dominant drivers have a strong influence on other units or series and their identification received growing attention in recent years. In an infinite VAR \cite{ChudikPesaran2013a} suggest to model the dominant driver as a common factor.  \cite{KapetaniosPesaranReese2020} propose a sequential multiple testing approach to identify drivers with pervasive effects in a large panel model. The underlying idea is to identify the drivers using their error variance, on which the multiple testing approach \citep{Bailey2020} is applied to. \cite{Parker2016} identify dominant drivers by analysing the residual variances of regressions of principal components on the time series and other principal components. \cite{PesaranYang2020} identify dominant drivers in production networks using a criterion similar to the exponent of cross-section dependence. \cite{Brownlees2021} define a dominant driver by the means of the column norms of the inverse of the covariance matrix and a selection criteria. Their criteria requires to invert the covariance matrix and is therefore only applicable to datasets with \(N<T\).

Our work extends the literature on inflation and the theoretical literature on the identification of dominant drivers. The results from the empirical application show that the inflation in the US and oil prices act as a dominating series. We do not find evidence that real GDP, equity prices, exchange rates and interest rates have a dominating effect on national inflation. 

The remaining part of this paper is structured as follows: the next section describes the theoretical background, followed by a discussion our \ddml. We provide evidence for our approach using Monte Carlo Simulations and discuss our findings of US and oil prices dominating national inflation rates.

The notation throughout this paper is as follows: matrices are in capital and bold, such as \(\boldsymbol{X}\), vectors are in lowercase letters and bold, \(\boldsymbol{x}\) and scalars are lowercase \(x\). Time indices are denoted by \(t=,1...,T\), unit indices by \(i=1,...,N\) and the number of variables are defined as \(k=1,...,K\). \(||\boldsymbol{X}_i||\) refers to the i-th column norm of matrix \(\boldsymbol{X}\).

\section{The \ddml approach}\label{sec:SeqEst}

This section defines the \ddmllong\ two step approach to identify dominant drivers in large panel models. In the first step we identify the links between the units and in the second step we select the dominant drivers.

In the first step, called the \textit{network selection step}, henceforth NSS, we recover the network structure using a graphical model and then estimate the concentration matrix. A graphical model links the estimation of a network in the form of nodes connected by edges and the structure of a covariance and concentration matrix, for a summary see \cite{Meinshausen2006,Sulaimanov2016,Friedman2008}. The link between the graphical model and the inverse of the covariance matrix is that if two series are independent they have a a zero partial correlation and are not in the same edge set.

We want to identify the set \(\Gamma(N_d)\) of \(N_d\) dominant drivers in the \(T \times N\) variable \(\boldsymbol{X}\). We define the entries of the dominant drivers as:

\begin{align}
		x_{i,t,d} &= u_{i,t},  & i &\in \Gamma(N_d) &
		\intertext{and of the non dominant units:}
    x_{i,t,nd} &=  \sum_{\substack{j\in \Gamma(N_{d})}}\beta_{i,j} x_{j,t}+ u_{i,t}, &  i &\not\in \Gamma(N_d) & \label{eq:eq1}
\end{align}
where \(\beta_{i,j}\) measure how much dominant driver \(j\) influences the non-dominant unit \(i\). A unit cannot influence itself, hence \(\beta_{i,i} = 0\). \(x_{i,t}\) and the iid error component \(u_{i,t}\) are can be serially correlated but stationary over time and can include common factors. The error component \(u_{i,t}\) can also be potentially heteroskedastic. We define the covariance matrix of \(\mathbf{X} = (\boldsymbol{x}_1,...,\boldsymbol{x}_N)'\) by \(\boldsymbol{\Sigma}\) and the concentration matrix is \(\boldsymbol{\kappa} = \boldsymbol{\Sigma}^{-1}\).

Following \cite{Meinshausen2006} \cite{Sulaimanov2016} the graphical model is recovered based on the optimisation problem:

\begin{align}
    \hat{\beta}(\lambda)_i = \arg \min_{\beta} \sum_{t=1}^T \left(x_{i,t} - \sum_{j=1, i\neq j}^N x_{j,t} \beta_j \right)^2 + \lambda_i \sum_{j=1}^N\psi_j |\beta_j| \label{eq:LassoNE}
\end{align}
where \(\lambda > 0\) is penalty level (or tuning parameter) and \(\psi_j\) is the penalty loading. To estimate Equation \eqref{eq:LassoNE} and select the penalty level and loading, we propose two estimator: the rigorous or plug-in LASSO \citep{Bickel2009a,Belloni2016,Chernozhukov2019,AhrensAitkenDitzenEtAl2020} and the adaptive LASSO \citep{Zou2006,Medeiros2016a}. The rigorous LASSO is a data driven method to select \(\lambda\). The penalty loading \(\psi_j\) is estimated and adjusted to the assumptions of the error variances and can account for clustered error variances \citep{Belloni2016}, heteroskedasticity and autocorrelated errors \citep{AhrensAitkenDitzenEtAl2020}. The adaptive LASSO is a two step method which allows simultaneous estimation and consistent variable selection by weighting the \(\ell_1\) penalty term. The penalty loading is obtained from an initial regression and the tuning parameter is obtained by cross-validation (CV) or information criteria such as the AIC, BIC, AICC.

\cite{Meinshausen2006} show that the inverse of the covariance can be estimated for each node or variable individually. This implies that the problem in Equation \eqref{eq:LassoNE} is repeated for each unit or variable which is influenced by the potential dominant driver. Depending on the data, the dominant driver can be a specific cross-sectional unit or a variable.

The solution to Equation \eqref{eq:LassoNE} yields the non-zero elements in each row of the concentration matrix, or in different words it informs which units influence the unit in question, unit \(i\). To construct the concentration matrix, the post-LASSO estimates \(\hat{\beta}_{i,j}\) for each cross-section are collected and the \(N\times N\) matrix \(\boldsymbol{\beta}\) constructed:

\begin{align}
\boldsymbol{\hat{\beta}} &= \left(\boldsymbol{\hat{\beta}}_1,...,\boldsymbol{\hat{\beta}}_N\right)' \\
\boldsymbol{\hat{\beta}}_i &= \left(\hat{\beta}_{i,1},...,\hat{\beta}_{i,i-1},0,\hat{\beta}_{i,i+1},...,\hat{\beta}_{i,N}\right)
\end{align}

Following  \cite{Sulaimanov2016} we construct the concentration matrix as:

\begin{align}
\mathbf{\hat{\kappa}} &= \mathbf{\hat{D}} \left(\mathbf{I}-\boldsymbol{\hat{\beta}}\right)\\ \label{eq:ConcMat}
\mathbf{D} &= diag(\hat{\sigma}_1,...,\hat{\sigma}_N)
\end{align}
where \(\boldsymbol{\beta}\) is a \(N\times N\) matrix of the estimated post-LASSO coefficients from \eqref{eq:LassoNE} and \(\mathbf{D}\) is a diagonal matrix with the inverse of the error variances of the i-th regression. The matrix \(\boldsymbol{\hat{\beta}}\) will be sparse and the sparseness will carry over to the concentration matrix \citep{Sulaimanov2016}. 

The second step, the \textit{dominant driver selection}, henceforth DDS, is based on \cite{Brownlees2021}. The authors show that the entries and therefore the column norm of the concentration matrix will be larger for dominant than non-dominant units.\footnote{The same applies to the row norm, for the remainder of this work we will use the column norm.}

The problem can be divided into two problems: the estimation of the number of dominant drivers and the identification of which units are dominant drivers. \cite{Brownlees2021} propose to use the eigenvalue ratio criterion from \cite{AhnHorenstein2013} to select the number of dominant drivers applied to ordered column norms of the concentration matrix. The number of dominant drivers is defined as:

\begin{align}
\hat{N}_d = \operatorname*{arg\,max}_{s=1,...,N-1} ||\hat{\kappa}_{(s)}||/||\hat{\kappa}_{(s+1)}|| \label{eq:BMest}
\end{align}
where \(\hat{\kappa}_{(s)}\) is the s-largest column norm of matrix \(\boldsymbol{\hat{\kappa}}\). All units with a larger column norm than column \(N_d\) are considered dominant drivers. \cite{Brownlees2021} show that their approach can be easily extended to common factors. The number of dominant drivers is a subset of all units, however no ratio is assumed.

The \textit{NSS} step requires the assumptions from \cite{Meinshausen2006} to ensure oracle properties of the estimator. The oracle properties imply that model selection and estimation of \(\boldsymbol{\beta}\) are unbiased and consistent. The properties are important because otherwise the second step, the DDS, will select falsely non dominant units as dominant drivers. In summary the assumptions for the oracle properties in \cite{Meinshausen2006} are that the graph is sparse, independence in the error components, correlations are bounded from below and neighbourhood stability. 
The aim of \ddml is to identify dominant drivers in high dimensional datasets where the number of cross-sections or variables is larger than the number of time periods. The dominant units are ordered in a block structure as in \cite{Brownlees2021}, however we explicitly allow for \(N>T\). The column norms of the dominant drivers are larger than a threshold and larger than those of non-dominant units. This is equivalent to a sparse concentration matrix, in which elements of non-dominant units are close to or exactly zero. The graphical model acts a thresholding method to select only the influential connections between units, which then imply a larger column norm in the concentration matrix.\footnote{Alternative methods for the estimation of the concentration matrix including thresholding the sample covariance matrix are discussed in \cite{Sulaimanov2016}.} The second assumption is that the sole source of dependence between units is via the dominant drivers and the residuals are cross-sectionally independent. While this assumption is restrictive, our simulations show that it can be relaxed to a certain degree. The last two assumptions are technical and ensure that the entries of the concentration matrix are not going to infinity and that there are no circle connections between two units via a third one. 


There are two notable challenges when applying the BM procedure to an estimated sparse concentration matrix. First the procedure can falsely select non-dominant units if the diagonal elements are very large in comparison to the off diagonal elements in a given column. The diagonal elements are the inverse of the residual variance. In combination with the LASSO estimator which minimizes the RSS, the residual variance can become relatively small making the diagonal element in the concentration matrix large in relation to the off diagonal elements. A second challenge is that the BM procedure does not take the number of non-zero elements, or connections, into account. In the extreme, the BM procedure can therefore select a unit with none or a small number connections to other units. To avoid this issue, we restrict the selection in the Monte Carlo Simulation and the empirical exercise.\footnote{\cite{KapetaniosPesaranReese2020} discuss this issue and call it the \textit{modified} BM procedure.} We also assume at least one dominant driver. The advantages of the BM procedure are besides the simple implementation the robustness to common factors, something which is confirmed in our Monte Carlo Simulations. 

Finally it should be noted that the the sample covariance matrix can be degenerated if \(N>T\). \cite{Bien2011} discuss this case assuming that the selected data is only a subset of the underlying data. The missing units are not connected to the sample. We can ignore those units in our approach since non-connected units will have no influence on the selection of the dominant drivers. Secondly we note that the matrices \(\boldsymbol{\hat{\beta}}\) and \(\boldsymbol{\hat{\kappa}}\) are not symmetric. \cite{Meinshausen2006} ensures symmetry by using the \textit{AND} criterion. Nodes \(i\) and \(j\) are connected if \(\beta_{i,j}\neq0\) and \(\beta_{j,i}\neq0\). Since we are interested in directed networks, \(\boldsymbol{\hat{\beta}}\) and \(\boldsymbol{\hat{\kappa}}\) are required to be non-symmetric. Therefore in the DDS step we can only use the column norm and not the row norm.\footnote{In a footnote BM state that the row norm can be used instead because of the symmetry of the concentration matrix.}

\section{Monte Carlo Simulation}

To show the Oracle properties of the proposed estimator, we employ a Monte Carlo Simulation. The simulations aims to shed light on three criteria: 1) model selection; that is if the individual LASSO estimators select the correct units; 2) if the number of dominant drivers is correctly estimated and 3) if the correct dominant drivers are selected. In total we are comparing 5 different specifications. Our data generating process follows \cite{KapetaniosPesaranReese2020}:

\begin{align}
    \boldsymbol{y}_{t,d} &= \boldsymbol{\mu}_d + \boldsymbol{\gamma}_d \boldsymbol{g}_t  + u_{d,t} \ , & i &= 1,...,N_d \label{eq:MC_dom} \\
    \boldsymbol{y}_{t,nd} &= \mu_{nd} + \boldsymbol{\beta} \boldsymbol{y}_{t,d} + \boldsymbol{\gamma}_{nd} \boldsymbol{g}_t  + u_{nd,t}\ , &i &= N_{d}+1,...,N \label{eq:MC_nondom}
\end{align}

where \(\boldsymbol{y}_{t,d}\) denotes a \(N_d \times 1\) vector of the \(N_d\) dominant drivers and \(y_{t,nd}\) a \(N_{nd}\times 1\) vector of the \(N_{nd}\) non dominant units. The fixed effects \(\boldsymbol{\mu}_d\) and \(\boldsymbol{\mu}_{nd}\) are drawn from a uniform distribution with \(IIDU(0,1)\).

\(\boldsymbol{\beta}\) is a \(N_{d} \times N_{nd}\) matrix and measures the impact of the dominant drivers on the non-dominant units and the individual elements are generated as:
\begin{align*}
    \beta_{ij} =
    \begin{cases}
    IIDU(0,1), & \text{if } i \leq \lfloor \left(N - N_d\right)^{\alpha}\rfloor \\
    0, & \text{otherwise}
    \end{cases}
\end{align*}

In the case of \(\alpha = 1\) the dominant drivers affect all non dominant units. In the case of \(\alpha<1\), only a subset is affected by the dominant drivers. For a discussion see \cite{KapetaniosPesaranReese2020}. The common factors \(f_t\) are uncorrelated across time and the loadings \(\boldsymbol{\gamma}_{d}\) and \(\boldsymbol{\gamma}_{nd}\) are drawn for each unit separately from a \(IIDU(0,1)\) distribution.\footnote{For more details see Section \ref{sec:DetMCSetting} in the Appendix.} The random noise \(u_{d,t}\) and \(u_{nd,t}\) are allowed to be correlated over time, measured by \(\rho_i\) and generated as a Gaussian process. The weak cross-sectional dependence in \(u_{d,t}\) and \(u_{nd,t}\) is measured by \(\rho_d\) and \(\rho_{nd}\). The dependence structure over time and space is varied between the different specifications.

\begin{table}[h]
    \centering
    \begin{tabular}{l c c c c c  c} \hline \hline
        Specification & \(\rho_d\) & \(\rho_{nd}\)  & \(\rho_i\)  & \(\alpha\) &  \(N_d\) \\ \hline
        (1) & 0 & 0 &0 & 1 &  5 \\
        (2) & 0 & 0 &0 & 0.5 &  5 \\
        (3) & 0 & 0 &IIDU(0.2,0.5) & 1 &  5 \\
        (4) & IIDU(0.2,0.5)  & 0.5 &0 & 1 &  5 \\
        (5) & 0 & 0 &0 & 1 &  \(f(N)\) \\\hline \hline
    \end{tabular}
    \caption{Monte Carlo Specifications. \(N\) and \(T\) are varied between 50, 100 and 150.  \(\rho_d\) and \(\rho_{nd}\) control the degree of cross-section dependence in the random noise \(u_{d,t}\) and \(u_{nd,t}\). Autocorrelation in the noise in present if \(\rho_i\neq0\). \(\alpha\) controls the strength of the dominant units, with \(\alpha = 1\) affecting all units and \(\alpha = 0.5\) affecting only a share. The number of dominant drivers \(N_d\) in Specification (5) are \(N_d=0.1N\), \(N_d=0.5N\) and \(N_d=0.9N\).}
    \label{tab:MC_specs}
\end{table}

Specification 1 is the simplest, with neither autocorrelation or dependence in the normally distributed (Gaussian) errors. The dominant driver affects all units. 
Specification 2 allows for weakly dependent dominant drivers by changing \(\alpha\) to \(0.5\). 
Specification 3 is an alternation of specification 1 and allows for autocorrelation in the errors. 
Specification 4 relaxes the cross-section independence assumption of the errors. 
Finally, Specification 5 is the same as specification 1 but the number of dominant drivers increases with the number of cross-sections.

The number of time periods and cross-sections varies between \(N\) and \(T=50,100,150\). The number of dominant drivers is fixed to \(5\) and the factors varies between 0, 1 and 5.\footnote{Additional simulations with 0 and 1 dominant factors, specifications with \(\chi^2\) distributed errors and different dependence structures are available in the appendix.} We present results for the HAC robust rigorous LASSO and the adaptive LASSO.\footnote{We also considered the elastic net LASSO in some preliminary simulations. Results were qualitatively worse than the adaptive or rigorous LASSO. We also acknowledge that other LASSO estimation methods such as the graphical LASSO \citep{Friedman2008} can be employed to select the model.} To select the hyperparamter \(\lambda\) of the adaptive LASSO we use the AIC, AICC or BIC criterion. The first stage loadings \(\hat{w}\) are calculated \(\hat{w}_j = 1/abs(\hat{\beta}_j)\) where \(\hat{\beta}_j\) are the from an univariate OLS regression of \(y_{i,t}\) on \(y_{j,t}\) \citep{Zou2006,Huang2008}.  We employ all estimators and estimate the \(N \times N\) matrix \(\boldsymbol{\beta}\). In the first step we use the LASSO estimators to select the non zero elements for each cross-section unit (row), then use post-LASSO OLS to estimate the coefficients in the matrix \(\boldsymbol{\beta}\) and finally construct the concentration matrix \(\boldsymbol{\hat{\kappa}}\) following Equation \eqref{eq:ConcMat}. We then calculate the column norms, order them by size and use the BM procedure based on Equation \eqref{eq:BMest}.

To assess if the LASSO estimators select the correct dominant drivers in the individual estimations we calculate the average number of non-zero elements in each column of \(\kappa\). Additionally we present the True Positive Rate (TPR), the False Positive Rate (FPR) and the False Discovery Rate (FPR) which are calculated as:

\begin{align}
TPR_{N} &=\frac{\underset{i\neq j}{\sum \sum }I(\hat{\kappa}_{ij}\neq 0,%
\text{ and }\kappa_{ij}\neq 0)}{\underset{i\neq j}{\sum \sum }I(\kappa
_{ij}\neq 0)}  \label{TPR} \\
FPR_{N} &=\frac{\underset{i\neq j}{\sum \sum }I(\hat{\kappa}_{ij}\neq 0,%
\text{ and }\kappa_{ij}=0)}{\underset{i\neq j}{\sum \sum }I(\kappa_{ij}=0)},
\label{FPR}\\
FDR_{N}&=\frac{\underset{i\neq j}{\sum \sum }I(\hat{\kappa}_{ij}\neq 0,%
\text{ and }\kappa_{ij}=0)}{\underset{i\neq j}{\sum \sum }I(\kappa_{ij}\neq
0)}.  \label{FDR}
\end{align}%

For the dominant drivers we perform the same analysis. We present the average number of estimated dominant drivers and assess if the correct ones are selected by comparing the TPR, FPR and FDR.

The simulations are done in R for the adaptive LASSO using \texttt{glmnet} \citep{Friedman2010} and repeated 1000 times. For the rigorous LASSO the Stata command \texttt{rlasso} \citep{Ahrens2020} is used with 100 repetitions.\footnote{We use a correction for the \cite{Brownlees2021} criterion. If the criterion selects as the largest growth the last possible growth rate, we use the 2nd largest growth rate. This in particular happens if the column norm for specific units turns to zero.}

\subsection{Results}
We start with analysing the \textit{NSS} step of Specification 1 with 5 dominant drivers. Table \ref{tab:tab_speclasso1_5_5} shows the results for the cases with no, one and five common factors. \(\hat{s}\) is the average of non zero column norms and should be equal to \(\left(N_d(N_d-1) + (N-N_d) N_d \right)/ N = [4.9,4.95,4.97]\), as shown in the last block called ``Oracle OLS''. The rigorous and adaptive LASSO both overselect the number of non-zero elements in the \(\boldsymbol{\beta}\) matrix. While the adaptive LASSO tends to improve with an increase in \(N\) and \(T\), the rigorous LASSO tends to select more non zero elements. Still the true positive rate is relatively small, indicating that the rigorous LASSO misses out many non-zero elements. However this is not necessarily a disadvantage because the identification of the dominant driver relies on the column norms of the concentration matrix and thus on the size of the estimated coefficients. As long as the falsely selected non dominant drivers obtain a smaller entry in \(\hat\kappa{i,j}\) than the dominant drivers, the column norms of the dominant drivers will be larger than those of the non-dominant units. An advantage of the rigorous LASSO is that it falsely selects an element in the concentration matrix less often than the adaptive LASSO (FPR). If the number of common factors is increased, the adaptive LASSO tends to select better than the rigorous LASSO. This can be seen by the general increase in the TPR.

Next we turn to the estimation of the dominant drivers. While the number of non-zero elements in a given column of \(\kappa\) gives an indication if a unit is dominant or not, the size of the post LASSO coefficient matters more. Table \ref{tab:tab_spec1_5_5} shows the estimated number of dominant drivers and if the correct units were select for specification 1 with 5 dominant drivers. In the case of no common factors, the approach using the rigorous LASSO and adaptive LASSO, independent of the selection criterion, estimate the number of dominant drivers precisely. This is especially the case for \(T=150\). The TPR is exactly or close to 100\% implying that all dominant drivers are correctly identified. A special case for the adaptive LASSO is if \(N=T\), in which the number of dominant drivers is underestimated. A reason for this might be the estimation of the initial loadings, which is done by OLS for each unit separately and relies on 50, respectively 100 observations.\footnote{We tested a specification with \(N=50,100,150\) and \(T=N-5\) and results behave better.} Noteworthy is that for small \(T\), the rigorous LASSO underestimates the number of dominant drivers. Both, the FPR and the FDR are small and converge to zero for both LASSO estimators. For the remainder of the paper we will focus on the results from the estimation of the dominant drivers.

Next we allow the dominant drivers to be weakly dominant, meaning a dominant driver affects only a subset of the non-dominant units. We set \(\alpha=0.5\), implying that only the first 6 (\(N=50\)), 9 (\(N=100\)) and 12 (\(N=150\)) non dominant units are affected. The results are displayed in Table \ref{tab:tab_spec2_5_5}. Again the adaptive LASSO underestimates the number of dominant drivers if \(N=T\). The estimated number of dominant drivers using the rigorous LASSO is slightly downward biased as well, however the bias decreases with \(N\) and \(T\) increasing. Interestingly the rigorous LASSO improves with an increase in the number of factors, while the adaptive LASSO does much worse in comparison to Specification 1. It selects a smaller number of dominant drivers and falsely identifies non dominant drivers as such. A reason for this is that the adaptive LASSO tends to select less non-zero elements in the \(\boldsymbol{\beta}\) matrix and therefore raises the chance to miss out dominant drivers respectively gives more weight to incorrectly selected non-dominant units. Both LASSO approaches outperform the BM criterion, even for the case of \(N<T\).

Specification 3 allows for autocorrelated errors. Results are similar to Specification 1, however distortions when using the adaptive LASSO in the case of \(N=T\) are less pronounced. As expected, both methods control well for autocorrelation in the errors. Noteworthy is that the BM procedure performs well for the case \(N<T\), but is affected when the number of dominant drivers is smaller than the number of common factors. 

So far we assumed strong cross-section independence in the random noise. The sole source of dependence were the dominant drivers or the common factors. To relax this assumption, Specification 4, Table \ref{tab:tab_spec4_5_5}, allows for weak dependence in the random noise components. The \ddml approach is still robust to weak dependence. However the TPR for the cases with more than one dominant driver shrinks, especially for the rigorous LASSO as it underestimates the number of dominant drivers. An increase in T mitigates the problem, but the bias remains.

As a final exercise we return to specification 1 but increase the number of dominant drivers as \(N_d = h N\) with \(h = [0.1,0.5,0.9]\). Table \ref{tab:tab_spec5_1_5} shows the results for \(N_d = 0.1 N\), implying that \(10\%\) of the cross-sections are dominant drivers. In the case of \(h=0.1\), all methods identify the number of dominant drivers well, but as before the adaptive LASSO underselects if the number of common factors is \(5\). Similarly, the rigorous LASSO does better if the number of common factors increases. Both estimator improve with an increase in \(T\). If the share increases, it is getting harder for the estimator to identify the correct units. A reason for this is that the cut-off point in the BM criterion is less defined. This is in particular the case if \(h=0.9\), implying that \(90\%\) of the units are dominant ones.  

In general the results show that the our proposed method reliably estimates the correct number of dominant drivers and identifies the correct drivers. In comparison to the criterion from \cite{Brownlees2021} our method can be applied to datasets with \(N>T\). Both LASSO estimator have their strength and weaknesses. While the rigorous LASSO is less affected by common factors, the adaptive LASSO tends to do better in the presence of weakly correlated errors. The case of 5 dominant drivers and 5 common factors is interesting with respect that the method based on the rigorous LASSO identifies the correct dominant drivers and the correct number, while the adaptive LASSO performs poorly. A possible reason might again be the first stage of the adaptive LASSO and difficulties differentiating the common factors and dominant drivers.

\section{Dominant Drivers in National Inflation Rates}\label{sec:EmpApp}

In this section we turn back to the question if national inflation rates are exposed to dominant drivers. We use the GVAR database \citep{MehdiRaissi2020} with quarterly observations from 1979Q2 to 2019Q4 (\(T=163\)) for \(N_g=33\) countries. The data is in first differences, standardized and demeaned on a country level. Taking first differences is necessary to remove potential non stationary, which ensures that the covariances are not time dependent and that the post-LASSO estimates are unbiased. Standardisation is required to ensure that the coefficients from the sequential regressions of \ddml are estimated in the same drivers.

The years covered in the GVAR dataset are different to \cite{Ciccarelli2010} and more in line with \cite{Levin2021,Rogoff2003} who cover the years from 1980 onwards. A key difference to \cite{Ciccarelli2010} is that we investigate the effect of global inflation on national inflation. If a variable country combination is selected as a dominant driver, it implies that it is an important driver for the inflation rate in many countries.

We will start by applying the BM procedure to the inflation series of the GVAR dataset. However the procedure can only be applied to a single series and dominant drivers of national inflation might be influenced by further covariates. We therefore apply the \ddml approach afterwards which overcomes the two limitations of the BM procedure.

\subsection{Number of Common Factors and BM Procedure} 
In a first step we estimate the number of common factors using the criteria from \cite{Bai2002} and \cite{AhnHorenstein2013}. The panel criteria from \cite{Bai2002} identifies between 4 and 5 common factors and the panel information criteria 1. Both estimator from \cite{AhnHorenstein2013} point to 1 common factor. The latter results are in line with the finding in \cite{Ciccarelli2010} who find one common factor. In addition testing for strong cross-section dependence \citep{PesaranCD2015} confirms the occurrence of strong cross-section dependence. \\

\begin{figure}[!h]
\centering
\begin{subfigure}{.5\textwidth}
  \centering
  \includegraphics[width=\textwidth]{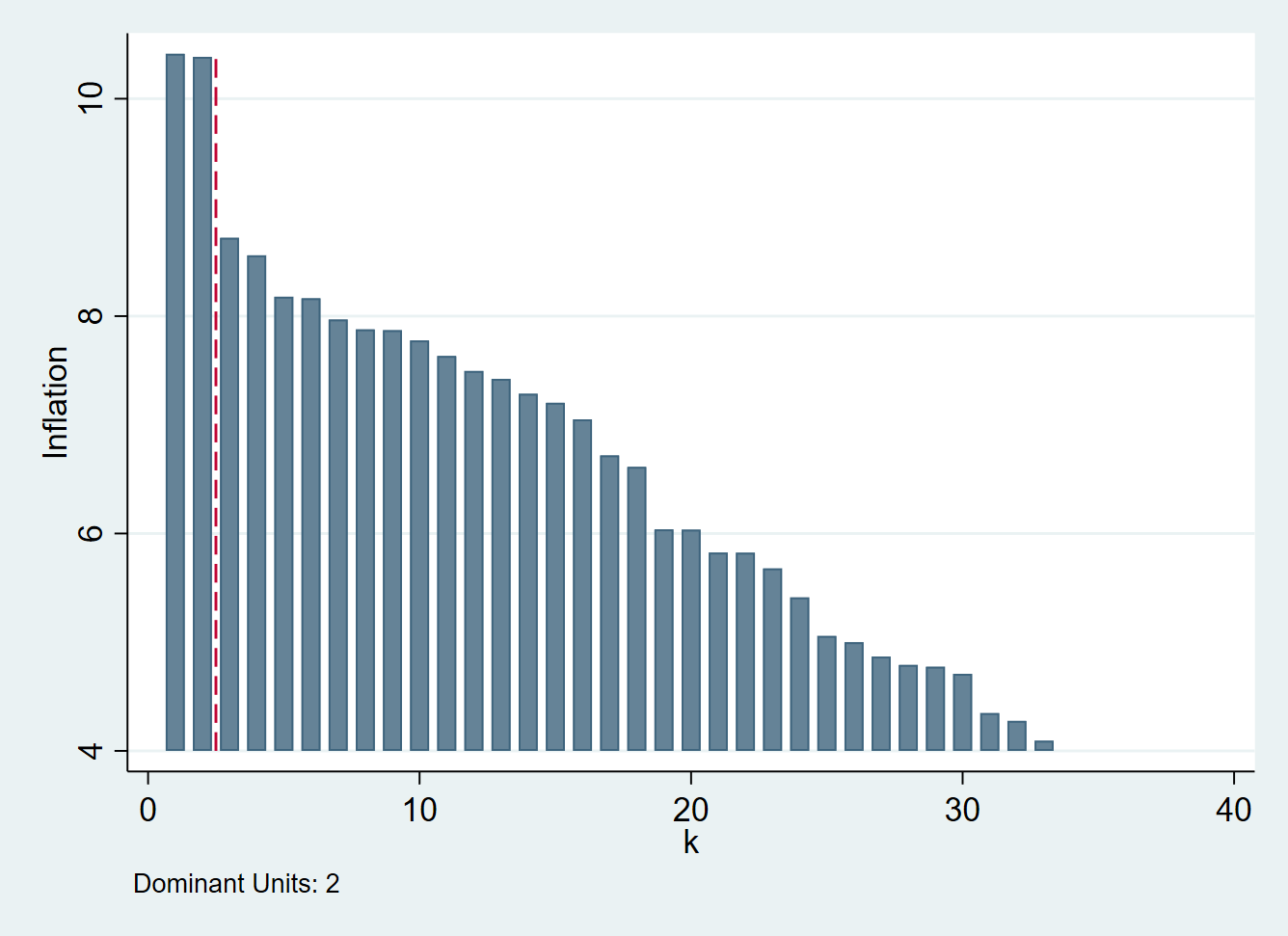}
  \caption{All Units}
\end{subfigure}%
\begin{subfigure}{.5\textwidth}
  \centering
  \includegraphics[width=\textwidth]{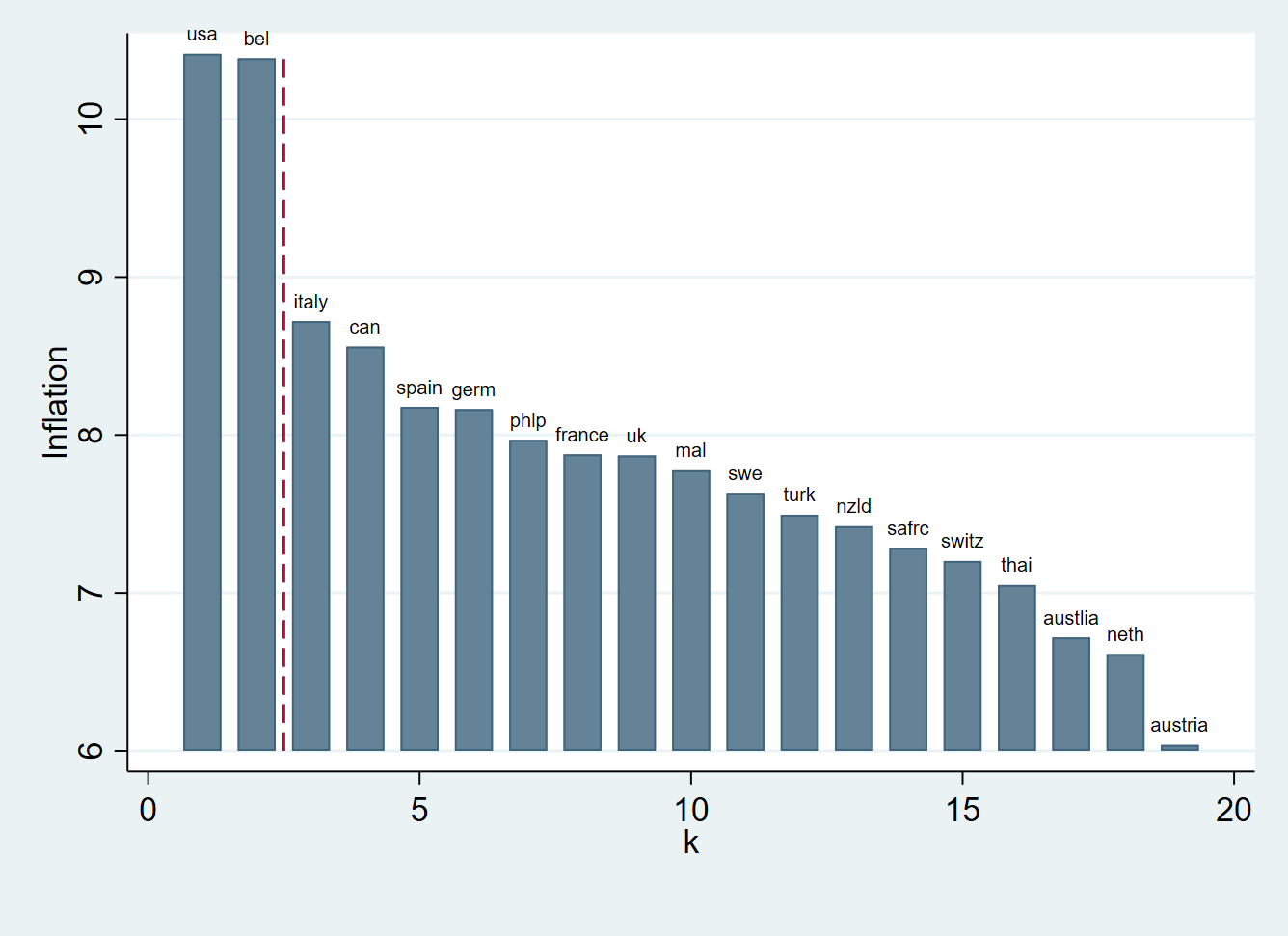}
  \caption{Largest 20}
\end{subfigure}
\caption{Column norms of the inverse of the sample covariance matrix of national inflation. Number of dominant drivers estimated by the BM procedure. See Table \ref{lab:ctry} for country name definitionsand section \ref{sec:EmpApp} for a detailed description.}
\label{fig:EmpAppBMC}
\end{figure}

The results imply an underlying common factor structure. To shed more light if the dependence structure is driven by common factors or dominant drivers, we apply the BM procedure next. Therefore we invert the sample covariance matrix to obtain the concentration matrix. Figure \ref{fig:EmpAppBMC} shows the results of the column norms. The dotted line indicates the dominant driver following the BM procedure. We find that the US and Belgium are dominant drivers in the national inflation series for the 33 countries. The two dominant drivers are connected to each other and all other drivers because the concentration matrix is non-sparse. While the US is somewhat expected to be a dominant driver, the finding that Belgium is a dominant drivers is surprising. However this is in line with \cite[Table 3]{Ciccarelli2010} who find that a large share of the detrended inflation variance of Belgium is explained by alternative measures of global inflation, pointing that Belgium's inflation rate is highly connected to others'.

A disadvantage of this approach is that the resulting concentration matrix picks up noise which can drive the determination of the dominant drivers. Secondly it is not possible to add any further covariates which might have an influence on national inflation and limiting the effects of the dominant drivers only on inflation. We therefore apply the \ddml approach next.

\subsection{Using the \ddml approach}

To allow other variables to have an effect on national inflation, we the \ddml approach as described in Section \ref{sec:SeqEst}. We model inflation as a function of inflation in other units (\(\mathbf{dp}_{-i}\)), real GDP (\(\mathbf{y}_{i}\)), real equity prices (\(\mathbf{ep}_{i}\)) and exchange rates (\(\mathbf{er}_{i}\)) and the nominal short run (\(\mathbf{r}_{i}\)) and long run interest rate (\(\mathbf{lr}_{i}\)). All variables are in first differences to remove potential non stationary and standardized for each country. Standardisation is necessary to ensure that all estimated coefficients are measured in the same units. An advantage of our approach is that if a variable has no influence on the inflation rate, the LASSO estimator will set the respective coefficient and thus influence to zero. In detail, we estimate the following model:

\begin{align}
\mathbf{Y}_{i,k} =&  \mathbf{X}_{-(i,k)} \boldsymbol{\beta}_{i,k} + \mathbf{e}_{i,k}, \ \ k = 1,..,K\\
\mathbf{X} =& \left(\mathbf{dp}, \mathbf{y}, \mathbf{ep}, \mathbf{eq}, \mathbf{lr}, \mathbf{r} \right) \label{eq:empAppX}\\
\boldsymbol{\tilde\beta}_k =& \left(\boldsymbol{\beta}_{1,k},...,\boldsymbol{\beta}_{N,k}\right)'\\	
\boldsymbol{\beta} =& \left(\boldsymbol{\tilde\beta}_1,...,\boldsymbol{\tilde\beta}_K\right)'
\end{align}

where \(\mathbf{Y}_{i,k}\) is the i-th element of the k-th variable of \(\mathbf{X}\). \(\mathbf{dp}, \mathbf{y}, \mathbf{dp}, \mathbf{ep}, \mathbf{eq}, \mathbf{lr}, \mathbf{r}\) are \(T \times N\) matrices and \(\boldsymbol{\beta}_{k,i}, k = 1,..,6\) are \(1 \otimes N\). The subscript \(-(i,k)\) denotes that the i-th element in the k-th variable of \(X\) is zero. \(\mathbf{e}_{i}\) is a \(T\times 1\) vector of random noise. \(\boldsymbol{\beta}\) is then a \(NK \times NK\) matrix which is used to calculate the concentration matrix. \(\boldsymbol{\tilde{\beta}}_k\) is a \(N\times NK\) matrix with the coefficients which measure the influence on the k-th variable. 

To estimate \(\boldsymbol{\beta}\), the following equation is estimated using the rigorous and adaptive LASSO:

\begin{align}
    \hat{\boldsymbol{\beta}}_i^{\lambda_{i}} = \arg\min_{\beta_i} \left(\mathbf{Y}_{i,k} - \boldsymbol{\beta}_i\mathbf{X}_{-(i,k)}\right)^2 + \lambda_i \sum_{j=1}^N \psi_i |\boldsymbol{\beta}_{-(i,k)}| \label{eq:EmpExLasso}
\end{align}

The concentration matrix \(\boldsymbol{\kappa}\) is calculated following Equation \eqref{eq:ConcMat}. Since we are only interested in the dominant drivers for inflation, that is in the first \(N\) off diagonal elements of \(\boldsymbol{\beta}\), we place further constraints on \(\boldsymbol{\kappa}\) for the application of the BM procedure. Let's denote \(\boldsymbol{\hat{\tilde{\beta}}^c}\) the constraint version of \(\boldsymbol{\hat{\tilde{\beta}}}\) and define it as:

\begin{align}
\boldsymbol{\hat{\tilde{\boldsymbol{\beta}}}^c} &= \begin{pmatrix}\multicolumn{2}{c}{\boldsymbol{\hat{\tilde\beta}}_1} \\ \mathbf{0} & I_{N(K-1)}\end{pmatrix}
\end{align}
Then the estimate of the concentration matrix \(\hat{\boldsymbol{\kappa}}\) becomes:
\begin{align}
\boldsymbol{\hat\kappa} &= \mathbf{\hat{D}}\left(I-\boldsymbol{\boldsymbol{\hat{\tilde{\boldsymbol{\beta}}}^c}}\right) \label{eq:EmpAppEstConcMat}\\
\mathbf{\hat{D}} &= diag(1/\hat\sigma_{i,k}^2)
\end{align}

The lower \((N-1)K\) rows of \(\boldsymbol{\hat{\tilde{\boldsymbol{\beta}}}^c}\) represent the effects of the covariates on all variables but inflation. Since we are only interested in inflation, this part is set to zero with the only exception of the diagonal. Non-zeros on the diagonal are required to ensure that the columns receive the equal weight for the BM procedure. The diagonal elements are the inverse of the residual variance of the regression representing the respective row. For example \(\hat\sigma_{1,1}\) is the variance of the residuals of inflation of Argentina on the variables selected in the NSS stage by the LASSO estimators.

For the selection of the dominant drivers, we use the modified version of the BM procedure. The selection is restricted to the \(N/2\) most connected units \citep{KapetaniosPesaranReese2020}. This has the advantage that we filter out units with very large values on the diagonal, but without any connection to other units.

\subsubsection{No Common Factors}
\paragraph{Rigorous LASSO}
First we employ the rigorous LASSO on Equation \eqref{eq:EmpExLasso} which allows for heteroskedasticity and autocorrelation of order 2.\footnote{We varied the bandwidth between 0, 1, 2, 4, and 8, to cover autocorrelation over several quarters. The results remain unchanged and are presented in the Online Appendix.} Column (1) in Table \eqref{tab:empAppStd} displays the results. We identify the inflation series of the US as the strongest and of Belgium as the second dominant drivers. Together the dominant drivers account for 57.58\% of the column norms and they influence in total 15 other units.\footnote{Diagonal elements are not counted in the column norm shares or the number of non-zero entries.} \ddml identifies 26 connections between units of which 13 are related to the two dominant drivers. Important to note is that the inflation rate in the US influences the inflation rate of 8 other countries, while the Belgium inflation rate influences 5 others. \\

\begin{sidewaystable}
\begin{tabular}{@{\extracolsep{4pt}}l ll ll ll ll @{}}\hline \hline
 & \multicolumn{2}{c}{(1)}& \multicolumn{2}{c}{(2)} & \multicolumn{2}{c}{(3)}& \multicolumn{2}{c}{(4)} \\
 & \multicolumn{4}{c}{Rigorous LASSO}& \multicolumn{4}{c}{Adaptive LASSO}  \\
\cline{2-5} \cline{6-9} 
Common Factors & &  No & & Yes & & No && Yes\\ 
\cline{2-3} \cline{4-5} \cline{6-7} \cline{8-9} 
Number of dom. Units & \multicolumn{8}{l}{} \\ 
1 & dp(bel) & 5 & dp(bel) & 5 & r(chl) & 5 & r(chl) & 5 \\
& & (27.38\%) & & (29.16\%) & & (23.90\%) & & (28.65\%) \\
2 & dp(usa) & 8 & dp(usa) & 7 & dp(usa) & 10 & dp(usa) & 10 \\
& & (30.19\%) & & (27.71\%) & & (19.64\%) & & (23.10\%) \\
3 & & & poil & 8 & & & & \\
& & & & (20.29\%) & & & & \\ \hline \multicolumn{9}{l}{Share columnnorms of dom. units.} \\
& & 57.58\% & & 77.17\% & & 43.54\% & & 51.76\% \\ \hline \multicolumn{9}{l}{No. Connections} \\
\multicolumn{2}{l}{\ \ Dominant} & 13 & & 20 & & 15 & & 15 \\
\multicolumn{2}{l}{\ \ Total} & 26 & & 26 & & 38 & & 33 \\

\hline\hline
\end{tabular}

\caption{Column norms of estimated concentration matrix based on Equation \eqref{eq:EmpAppEstConcMat}. The diagonal elements for calculation of column norm shares and number of connections are removed. Column norm shares for dominant drivers are in parenthesis. \textit{dp()} denotes national inflation rates, \textit{r()} national short run interest rates and \textit{poil} oil prices. Abbreviations for countries are United States (USA), Belgium (bel) and Chile (chl). See Table \ref{lab:ctry} for country name definitions and section \ref{sec:EmpApp} for a detailed description.}
\label{tab:empAppStd}
\end{sidewaystable}

\begin{figure}[!h]
\centering
\begin{subfigure}{.5\textwidth}
  \centering
  \includegraphics[width=\textwidth]{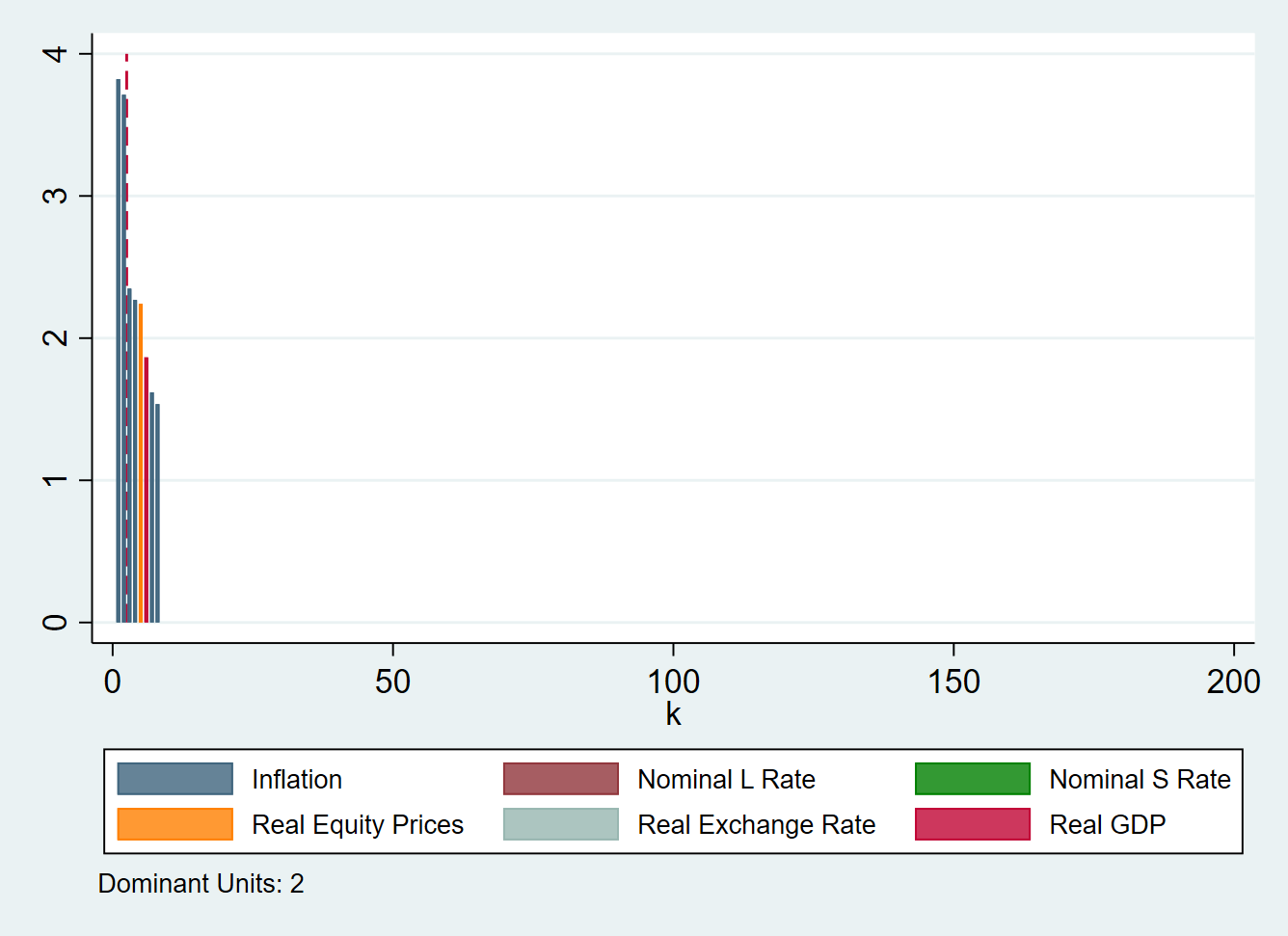}
  \caption{All Column Norms}
\end{subfigure}%
\begin{subfigure}{.5\textwidth}
  \centering
  \includegraphics[width=\textwidth]{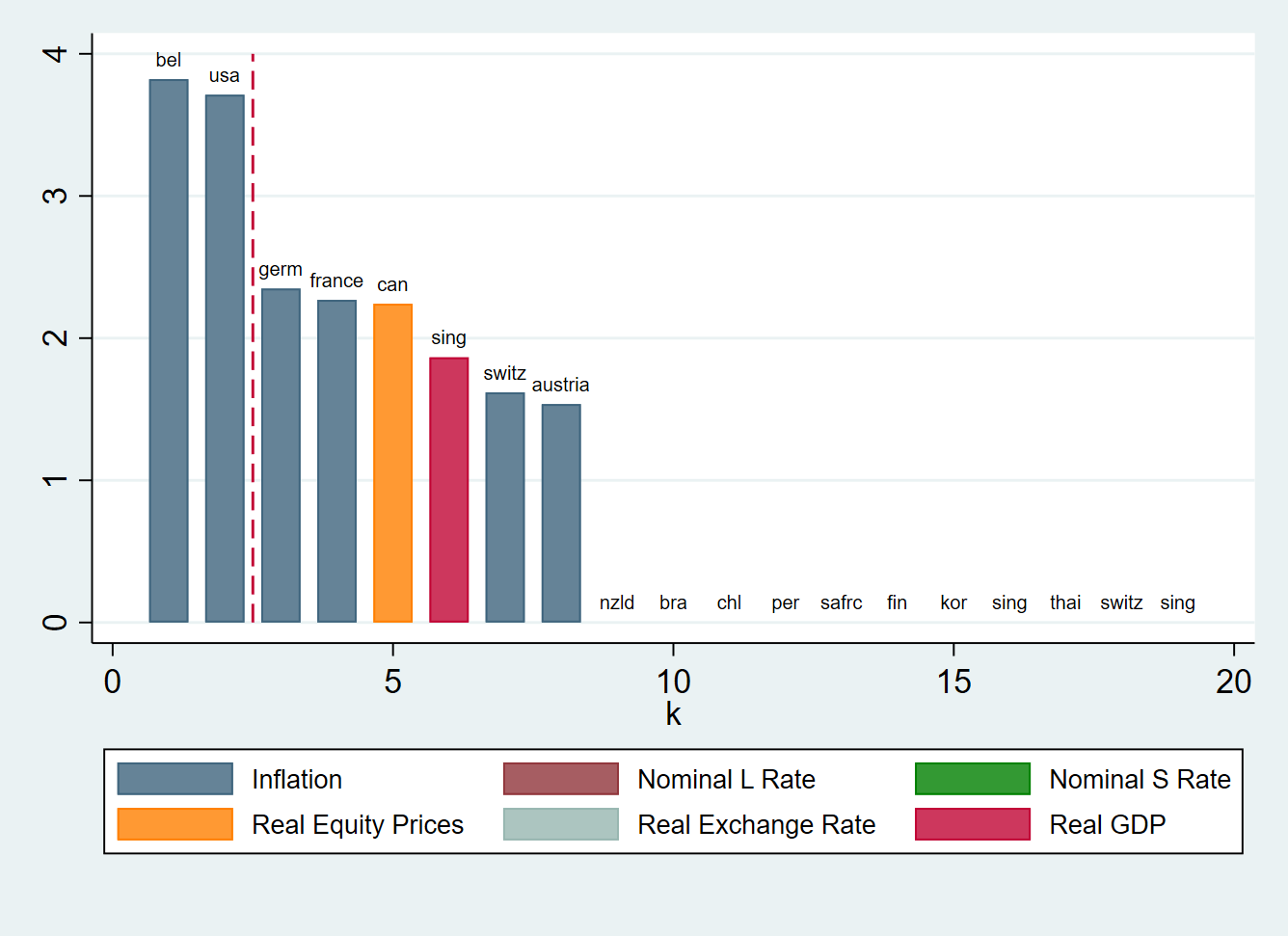}
  \caption{Largest 20 Column Norms}
\end{subfigure}
\hfill
\begin{subfigure}{\textwidth}
\centering
\includegraphics[width=\columnwidth]{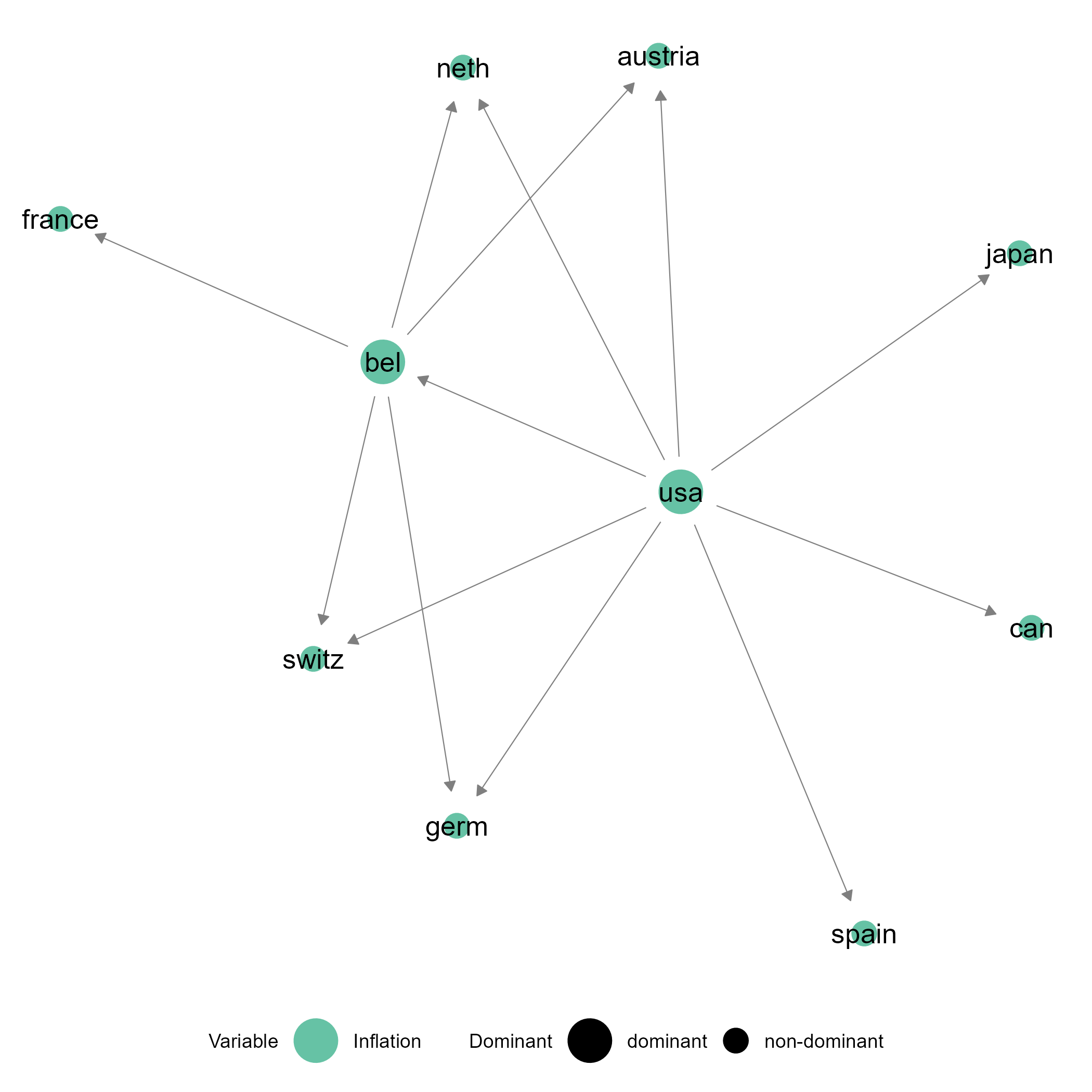}%
\caption{Network Graph rigorous LASSO}%
\label{fig:NGRlasso}%
\end{subfigure}
\caption{Column Norms and Network Graph for dominant drivers using rigorous LASSO in the NSS step. Norms to the left of the red dotted line are dominant drivers in Panel (a) and (b). See Table \ref{lab:ctry} for country name definitions and section \ref{sec:EmpApp} for a detailed description.
}
\label{fig:EmpAppRlasso2}
\end{figure}

An alternative method to display the results is to look at the column norms directly. Panel (a) of Figure \ref{fig:EmpAppRlasso2} shows the column norms across all country and variable combinations, Panel (b) only the largest 20. Each bar represents a country and a variable. For example the largest bar in Figure \ref{fig:EmpAppRlasso2} is the column norm of the US inflation. All country-variable combinations to the left of the dashed red line are dominant drivers. \\

Panel (a) shows the distribution of all column norms. We note that inflation has six non-zero column norms, real GDP and real equity prices have one non-zero column norm each. In Panel (b) it becomes evident that the growth rate from the US to the German interest rate is largest and therefore marks the border between dominant and non dominant units. 

The lower panel of the figure displays a network graph of the dominant drivers. It is noteworthy that the US influences Belgium but not the other way around. Belgium is a dominant driver for only European countries, while the US influences most of western Europe and Japan and Canada. The effect of Belgium is somewhat surprising. Besides the explanation discussed in the previous section, it is possible that the influence of Eurozone inflation on the national inflation rates not only within the Eurozone but outside of it are picked up by the Belgium inflation. Our finding is in line with argument in \cite{Bataa2013} that national inflation in Eurozone countries moves together and Belgium is a proxy for it. Further it extends the finding in \cite{Billo2016} that US leads the Eurozone cycle to inflation. \\

\paragraph{Adaptive LASSO}

Next we employ the adaptive LASSO as the selection method of the NSS stage.\footnote{We further restrict the selection of the dominant drivers such that the diagonal is less than 50\% of the absolute sum.} The adaptive weights origin from a univariate regression, similar to the Monte Carlo simulation.

\begin{figure}[!h]
\centering
\begin{subfigure}{.5\textwidth}
  \centering
  \includegraphics[width=\textwidth]{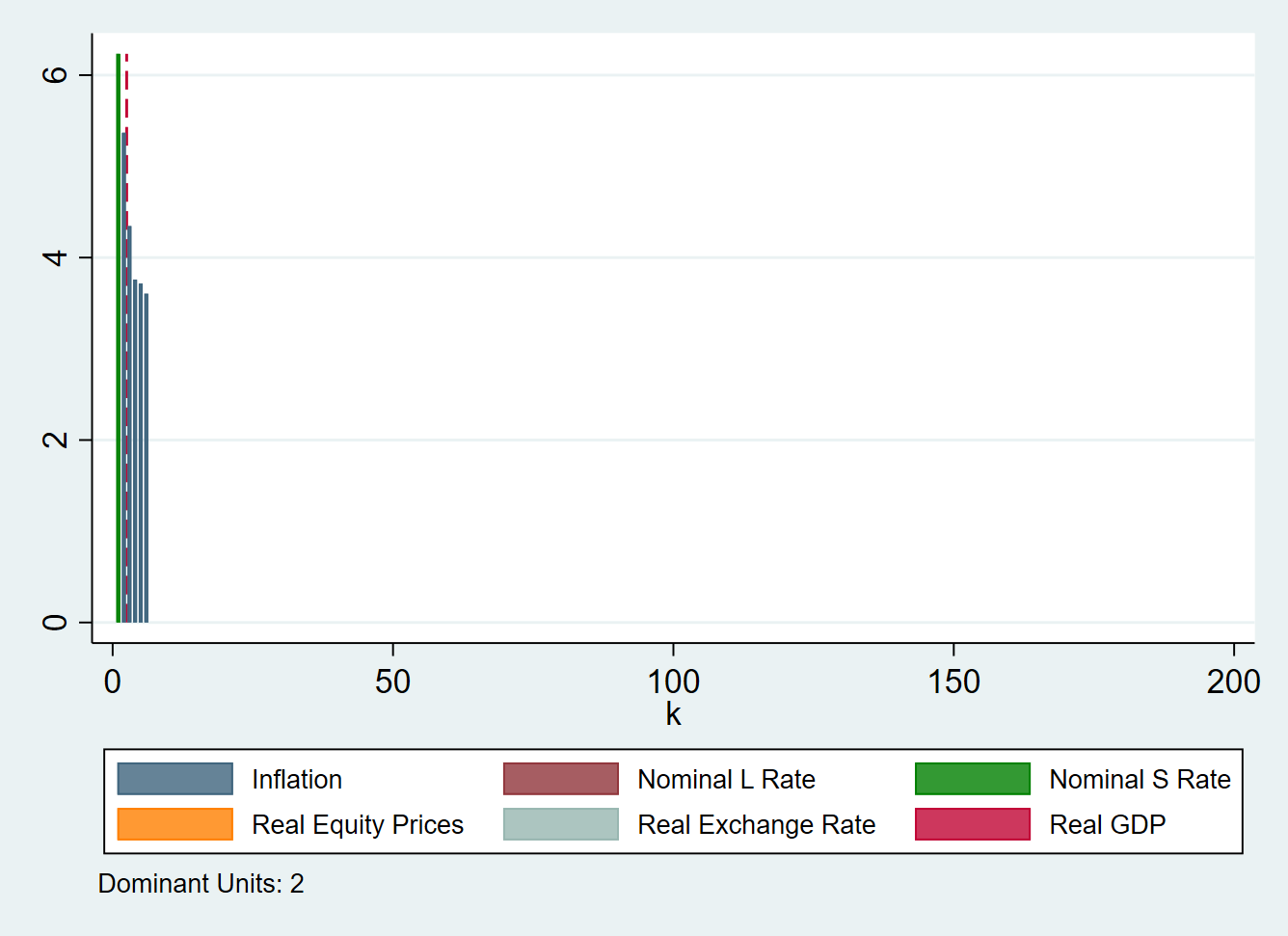}
  \caption{All Column Norms}
\end{subfigure}%
\begin{subfigure}{.5\textwidth}
  \centering
  \includegraphics[width=\textwidth]{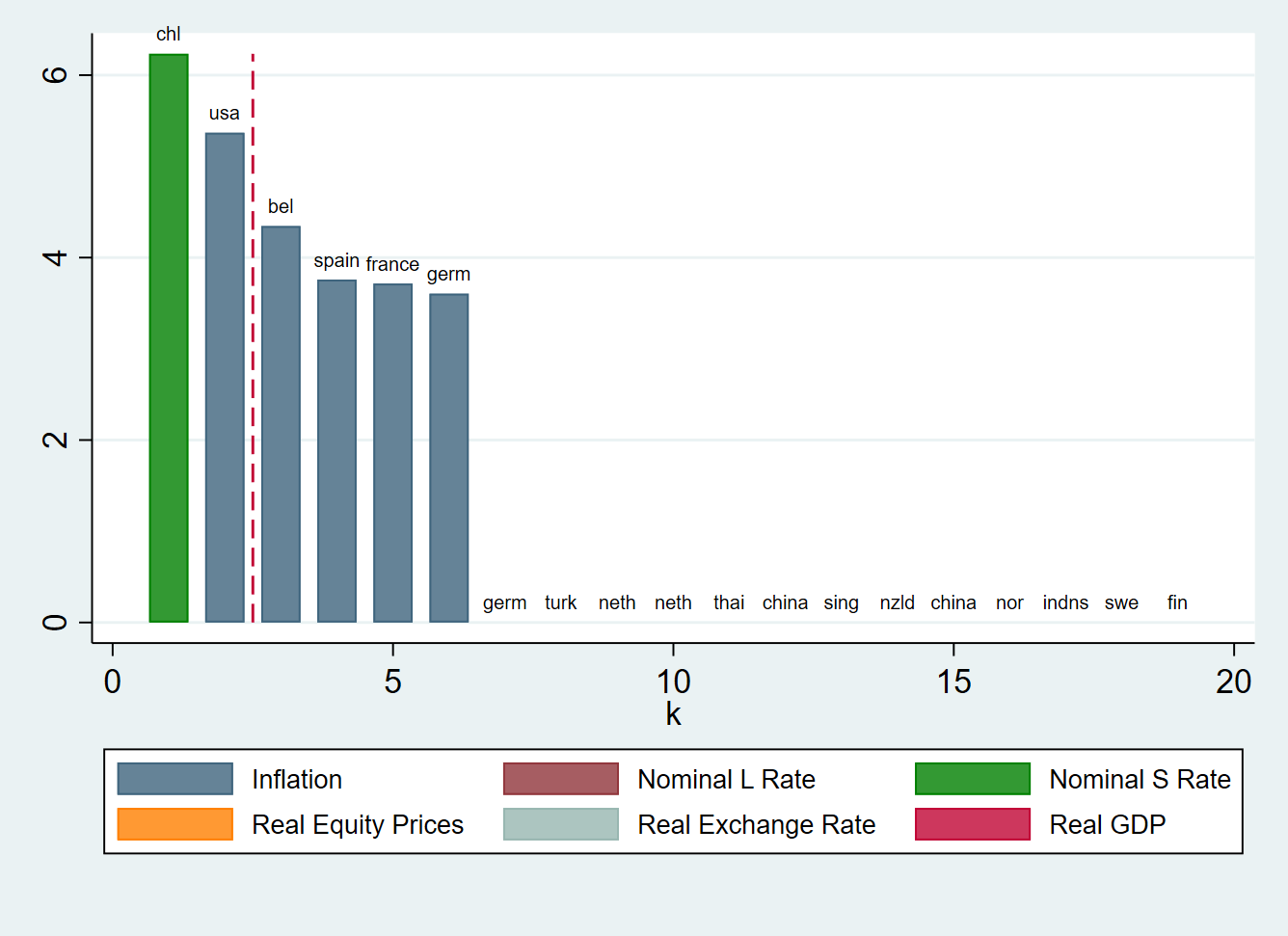}
  \caption{Largest 20 Column Norms}
\end{subfigure}
\hfill
\begin{subfigure}{\textwidth}
\centering
\includegraphics[width=\columnwidth]{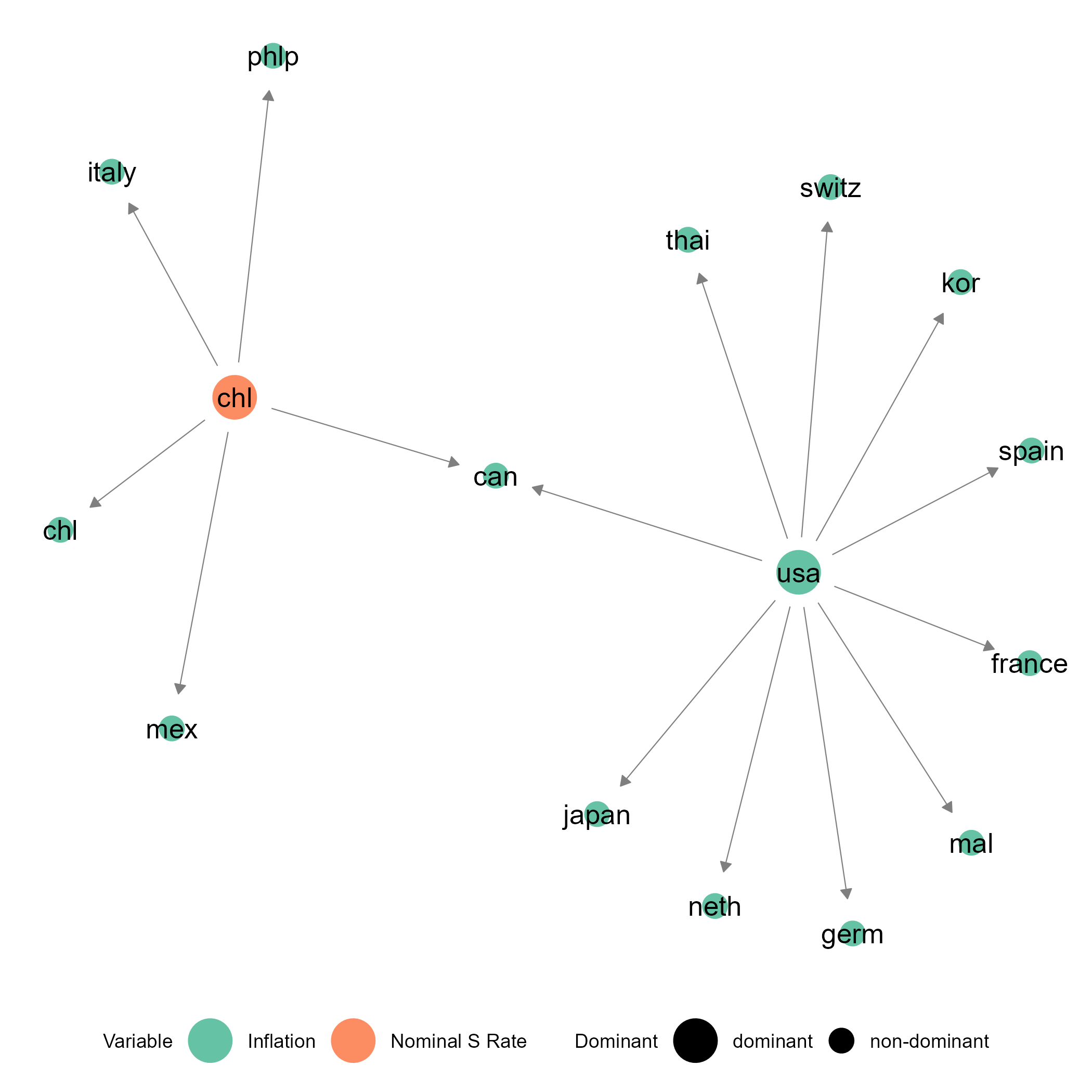}%
\caption{Network Graph adaptive LASSO}%
\end{subfigure}
\caption{Column Norms and Network Graph for dominant drivers using adaptive LASSO in the NSS step. Then penalty factor is selected by the BIC criterion. Univariate OLS is used in the caculation of the weights. Norms to the left of the red dotted line are dominant drivers in Panel (a) and (b). See Table \ref{lab:ctry} for country name definitions and section \ref{sec:EmpApp} for a detailed description.}
\label{fig:EmpAppAdalasso}
\end{figure}

The third column in Table \ref{tab:empAppStd} shows the results for this approach. Two dominant drivers are identified, the short run interest rate in Chile and as in the case of the rigorous LASSO the inflation rate in the US. While the US influences a larger number of drivers, the influence of the Chilean short run interest rate is stronger. The number of identified connections is much larger than in the case of the rigorous LASSO. 
Figure \ref{fig:EmpAppAdalasso} shows that the inflation series are again picked up most often influencing other inflation series. Panel (b) shows that inflation in Belgium is again influencing strongly other inflation rates, but it is not picked as a dominant driver. \\

Turning to the network graph in Figure \ref{fig:EmpAppAdalasso} it becomes evident that the short run rate in Chile not only influences the inflation rate in Chile but other countries as well. Among the main drivers for the column norm of the short run rate of Chile is however the affect on the inflation in Chile. The US has a similar widespread influence on inflation rates again. Together with the short run rate of Chile, the two dominant drivers influence 15 out of the 38 national inflation series.

Our results are in contrast to \cite{Ciccarelli2010} who find that no country is leading global inflation. Our results strongly suggest that the United States are a dominant factor for national inflation. Depending on the model selection method, Belgium respectively the short run interest rate in Chile are selected as dominant drivers.

\clearpage
\subsubsection{Observed Common Factors}

It is likely that inflation is not only driven by specific countries, but by other global factors. Examples would be commodity prices, see \cite{AasteveitBjornlandThorsrud2015}. To investigate the effect of such observed common factor, we add the prices of oil (\(p_{oil}\)), metals (\(p_{met}\)) and materials (\(p_{mat}\)) to the set of variables in matrix \(\mathbf{X}\) in Equation \eqref{eq:empAppX}:

\begin{align}
\mathbf{X} =& \left(\mathbf{dp}, \mathbf{y}, \mathbf{ep}, \mathbf{eq}, \mathbf{lr}, \mathbf{r}, \mathbf{p} \right) \\
\mathbf{p} =& \left(\mathbf{p}_{oil}, \mathbf{p}_{mat},\mathbf{p}_{met}\right)
\end{align}

\begin{figure}[!h]
\centering
\begin{subfigure}[t]{.5\textwidth}
  \centering
  \includegraphics[width=\textwidth]{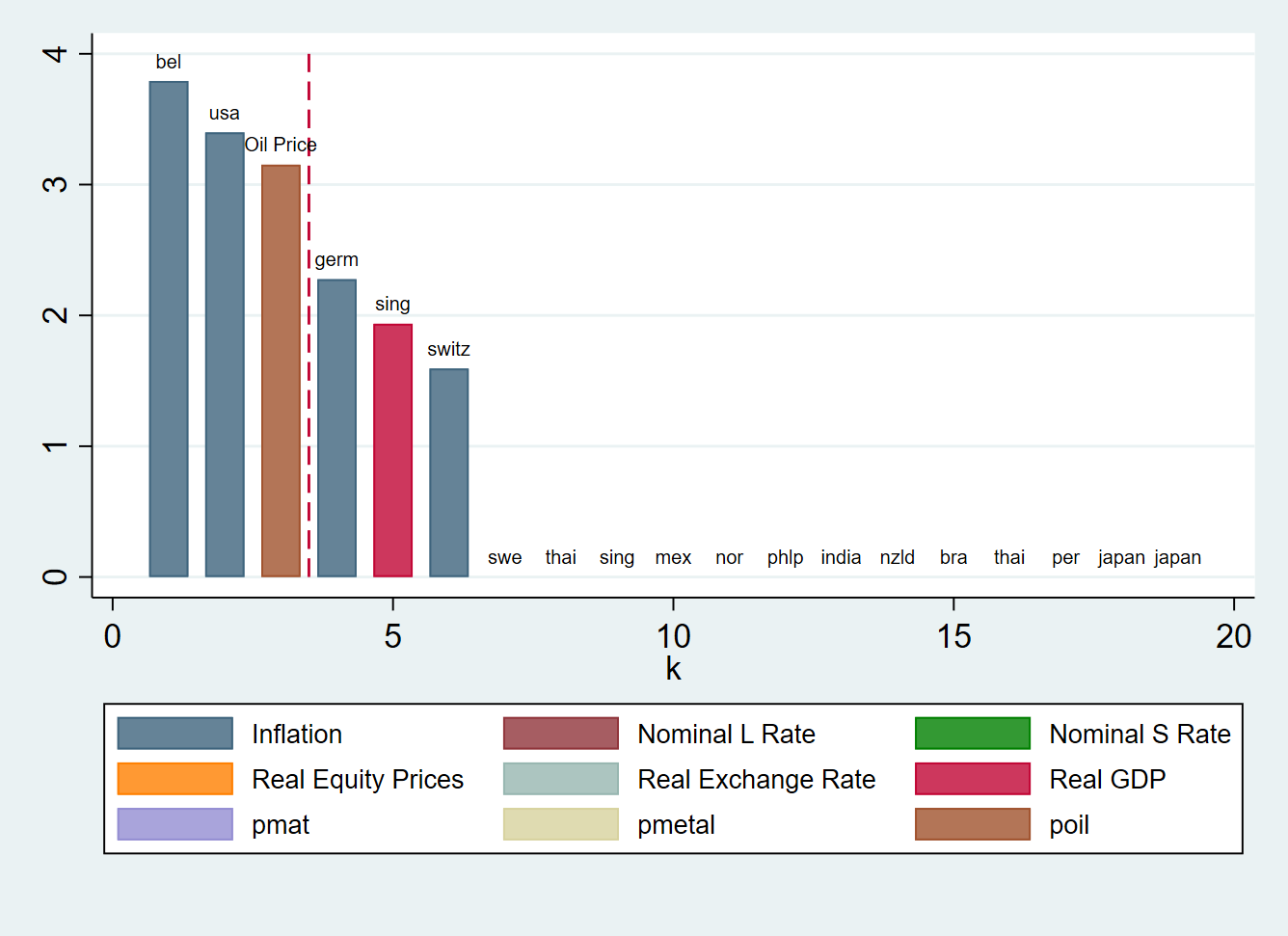}
  \caption{Column Norms from Rigorous LASSO}
\end{subfigure}%
\begin{subfigure}[t]{.5\textwidth}
  \centering
  \includegraphics[width=\textwidth]{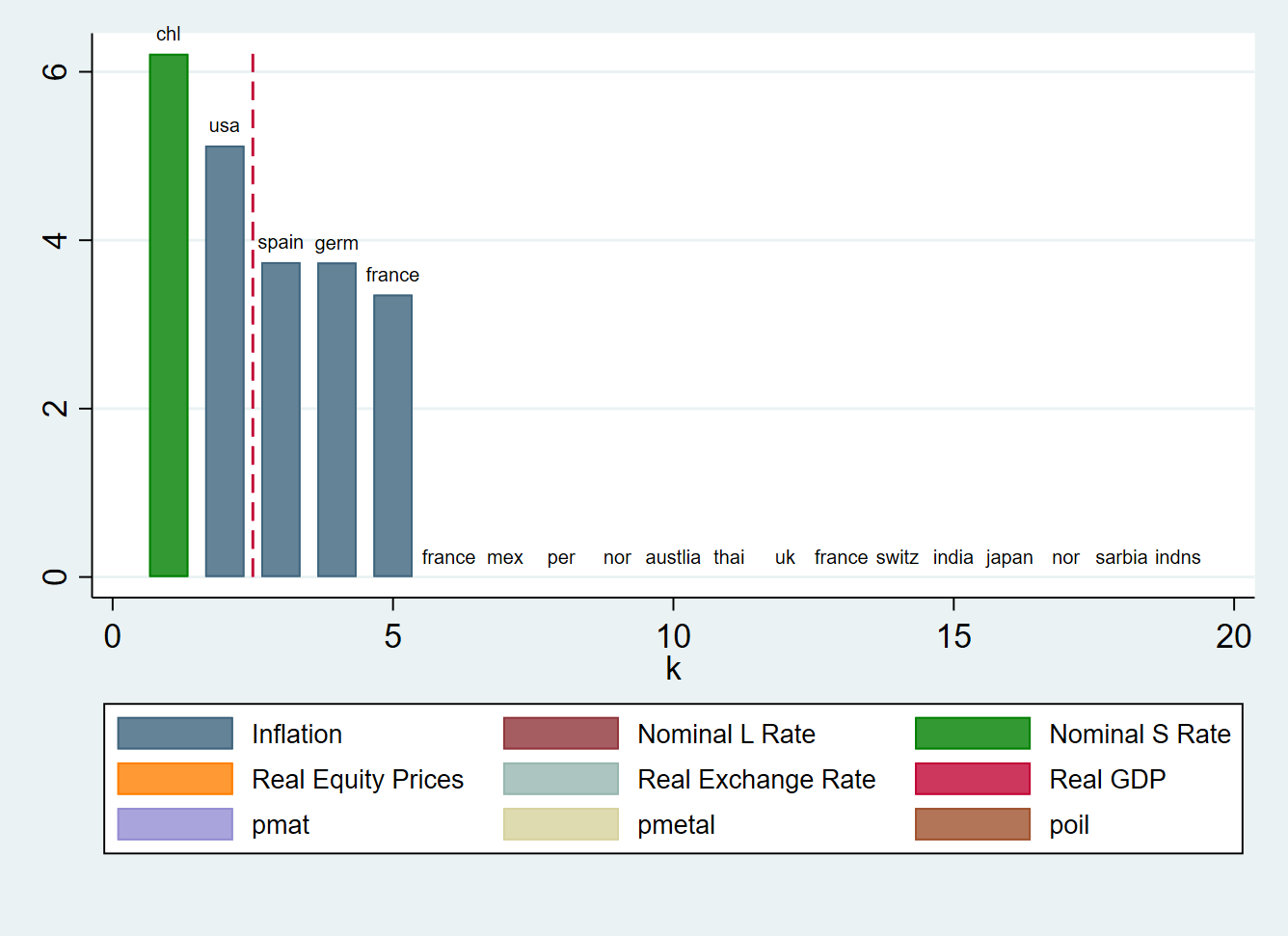}
  \caption{Column Norms from Adaptive LASSO (BIC)}
\end{subfigure}
\hfill
\begin{subfigure}[t]{\textwidth}
\centering
\includegraphics[width=\columnwidth]{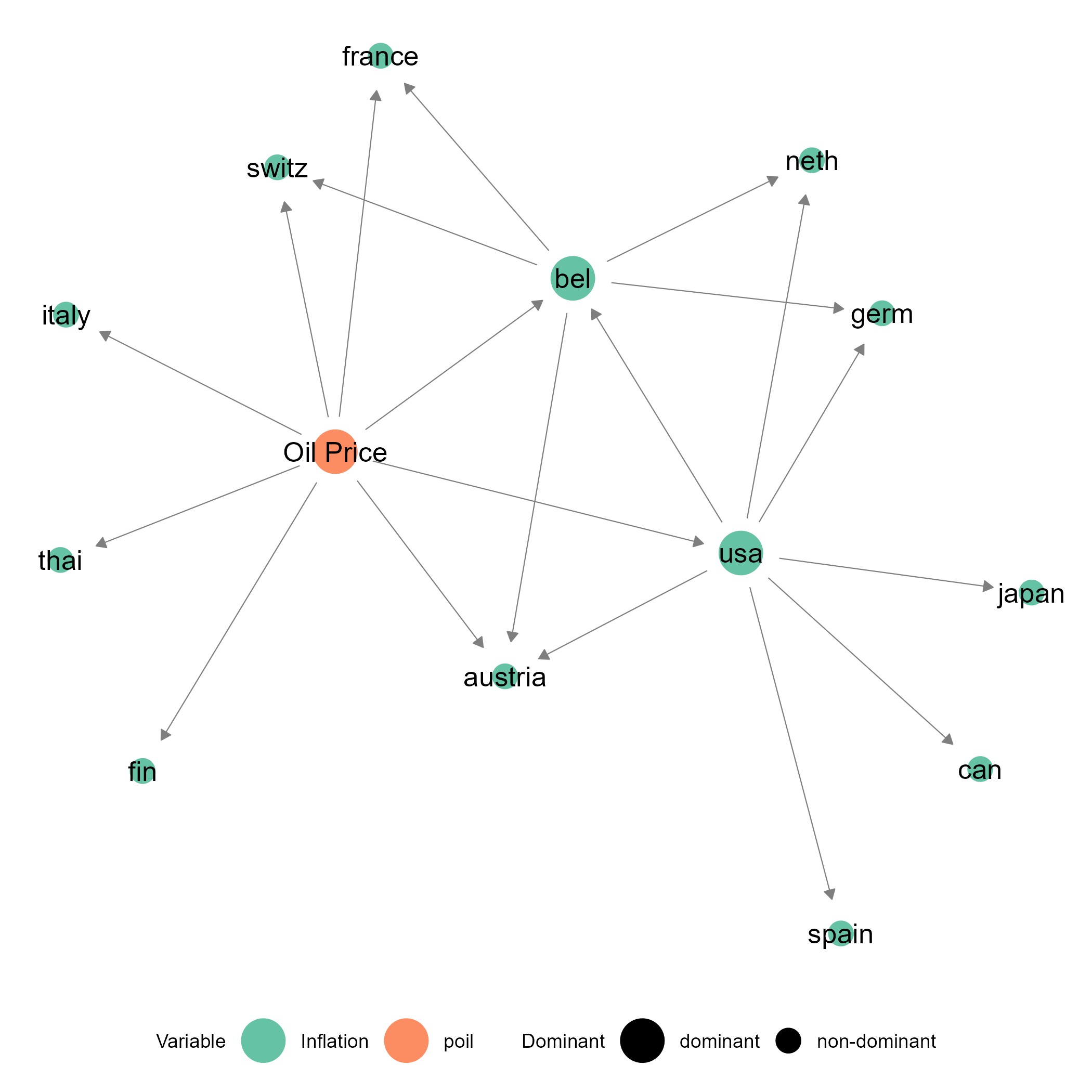}%
\caption{Network Graph Rigorous LASSO}%
\end{subfigure}
\caption{Commodity prices added as covariates.\\
Column Norms for the rigorous and adaptive LASSO. Norms to the left of the red dotted line are dominant drivers in Panel (a) and (b). See Table \ref{lab:ctry} for country name definitions, section \ref{sec:EmpApp} for a detailed description and note from Figures \ref{fig:EmpAppRlasso2} and \ref{fig:EmpAppAdalasso}.}
\label{fig:EmpAppObsCF}
\end{figure}

The commodity prices are the same across the 33 countries and they are allowed to be selected as a dominant driver. 
Results are presented in column (2) and (4) of Table \ref{tab:empAppStd} and in Figure \ref{fig:EmpAppObsCF}. A comparison between Column (3) and (4) reveals that the adaptive LASSO is not influenced by the additional observed common factors and the results remain similar. We therefore turn directly to the rigorous LASSO. Again the US and Belgium are selected as dominant drivers, however oil prices are selected as a third a dominant driver. In fact, they account for 20.3\% of the column norms. The network structure of the two other drivers remain almost unchanged. The US is not directly connected to Switzerland any longer but the shares of the column norm remain stable. As shown in the Monte Carlo simulation, not accounting for common factors does not worsen the identification of dominant drivers using the rigorous LASSO as a selection method.\\

The network graph in Figure \ref{fig:EmpAppObsCF} reveals that inflation in Austria, Finland, France, Italy, Switzerland and Thailand is influenced by the global oil prices. Oil prices are also influencing the two other dominant drivers, Belgium and the US, emphasising further their importance. 

Our results so far show that the US and oil prices play a crucial role for national inflation rates. In line with the results in \cite{Ciccarelli2010} is that some countries are sheltered from the dominant drivers. Our analysis shows that the inflation hardly spills over into countries such as China, India or Norway. Implying that for those countries other factors than the ones covered in our empirical application play an important role.

\clearpage
\subsection{Unobserved Common Factors}

We further account for unobserved common factors.\footnote{Detailed results are available in the Online Appendix.} We approximate potential unobserved common factors using principal components (PCA) or cross-section averages. Both methods are well established in the literature to account for unobserved common factors \citep{Pesaran2006,Bai2009}. In the previous section we found up to 4 common factors. Assuming that some of those are dominant drivers, we add the first 3 principal components (PCA). Separately the cross-section averages (CSA) of all variables are added to the model. The PCA and CSA are added in the same way as the observed common factors in the previous section.

The findings are similar to the case of adding observed common factors. The nodelasso using the adaptive LASSO identifies Chile as the sole dominant drivers. The rigorous LASSO only identifies the first principal component respectively the cross-section average of inflation as a dominant driver. This implies that the approximation of the unobserved common factors using accounts for too much of the variation and overlays the network structure. The result hints that the strong approximation of the common factor overlays weaker dependence structure, a similar finding than in \cite{Juodis2022}. Furthermore the long run interest rate in Germany is selected as a dominant driver if the bandwidth is equal to two years and it influences the inflation rate in Switzerland and the United Kingdom.

\section{Conclusion}
This paper combines the approach by \cite{Meinshausen2006,Sulaimanov2016} to estimate the inverse of a covariance matrix using time dependent data. It then identifies dominant drivers using the procedure from \cite{Brownlees2021}. Monte Carlo simulations show that \ddml identifies the correct units as dominant drivers in a panel. We then apply the method to the GVAR dataset and find evidence that inflation in the US and oil prices are dominant factors for national inflation rates. Our results are informative about potential spillovers into national inflation. It can also be used to improve forecasts using a  approach as in \cite{Bjornland2017}. 

While \ddml is simple, it has several limitations. First of all it is computationally expensive, in particular for a large number of cross-sections, respectively variables. Secondly as criticized by \cite{Yuan2007,Banerjee2008,Friedman2008} the approach by \cite{Meinshausen2006} is only an approximation to the problem. An extension in the spirit of this paper would be to use graphical LASSO \citep{Friedman2008} for time dependent data. The selection of an oracle estimator for the NSS step is crucial. It the selected estimator fails, the sequential estimator will fail as well. In our current setting higher order spatial effects and time varying dominant drivers can also not be considered and are left for future research.

\bibliographystyle{apsr2}
\bibliography{DR_DomUnits.bib}

\clearpage
\appendix
\section{Detailed Monte Carlo Settings}
\label{sec:DetMCSetting}
The loadings of the common factors \(f_t\) are generated as:
\begin{align}
   \boldsymbol{\gamma}_{d} &= \left(\gamma_{1},..,\gamma_{N_{d}}\right) \\
   \boldsymbol{\gamma}_{nd} &= \left(\gamma_{N_{d}+1},..,\gamma_{N}\right) \\
   \gamma_{i} &\sim IIDU(0,1) , \ i = 1,...,N
\end{align}

The unobserved common factors are generated as

\begin{align}
  \boldsymbol{g}_t &= \boldsymbol{R}_g^{1/2}\left(g_t-2\tau_k\right)/2\\
  g_t &\sim IID\chi^2(2) \\
  \boldsymbol{R}_g &= \left(1-\rho_g\right) \boldsymbol{I}_{m_{k}} + \rho_g \tau_k \tau_k'\\
  \rho_g &= IIDU(0.2,0.8) 
\end{align}

with \(\tau_k = (1,...,1)'\). 

The random noise \(u_{d,t}\) and \(u_{nd,t}\) are allowed to be correlated over time and generated as a Gaussian process:

\begin{align*}
   \boldsymbol{u}_{d,t} &= \left(1-\rho_i\right)\boldsymbol{R}^{1/2}_d \left(\boldsymbol{u}*_{d,t} - 2 \tau_{N_{d}}\right)/2 + \rho_i \boldsymbol{u}_{d,t-1}\\
    \boldsymbol{R}_d &= \left(1-\rho_d\right) \boldsymbol{I}_{N_{d}} + \rho_d \tau_{N_{d}} \tau_{N_{d}}'\\
    \boldsymbol{u}^*_{d,t} &\sim IIDN(0,1) 
\end{align*}
and for the non dominant drivers:

\begin{align*}
    u_{nd,it} &= \rho_i u_{nd,it-1} + \left(1 - \rho_i^2\right)^{1/2} \epsilon_{it}\\
    \boldsymbol{\epsilon}_t &= \left(\epsilon_{N_{nd}t},...,\epsilon_{Nt}\right) = \Sigma^{1/2}\boldsymbol{R}_{nd}^{1/2}\boldsymbol{\xi}_t\\
    \Sigma &= diag\left(\sigma_11,..,\sigma_{NN}\right)\\
    \sigma_{ii} &= IID\chi^2(2)/4 + 0.5\\
    \xi_t &= IIDN(0,1)\\
    \boldsymbol{R} &= \begin{pmatrix}
    1& \rho_{nd} &  \rho_{nd}^2 & \hdots &  \rho^{n_{nd}-1}_{nd} \\
    \rho_{nd} & 1 &  \rho_{nd}&   \hdots &  \rho^{n_{nd}-2}_{nd} \\
    \vdots & & & & \vdots \\
    \rho^{n_{nd}-1}_{nd} & \rho^{n_{nd}-2}_{nd} & \rho^{n_{nd}-3}_{nd} & & \hdots 1
    \end{pmatrix}
\end{align*}

\section{Monte Carlo Results}

\clearpage
 \begin{table}[h]     \scriptsize     \centering     \begin{tabular}{@{\extracolsep{4pt}} c c c c c c c c c c c c c}     \hline \hline  & \multicolumn{3}{c}{\(\hat{s}\)} & \multicolumn{3}{c}{TPR} & \multicolumn{3}{c}{FPR} & \multicolumn{3}{c}{FDR} \\ \cline{2-4} \cline{5-7} \cline{8-10} \cline{11-13} N/T & 50 &100 &150 & 50 &100 &150 & 50 &100 &150 & 50 &100 &150 \\ \hline \hline \multicolumn{13}{l}{Number of Factors: 0} \\ \hline \hline \multicolumn{13}{l}{rigorous LASSO (HAC robust)} \\ \hline 
50 & 46.08 & 51.73 & 55.29 & 15.37 & 25.59 & 30.81 & 8.25 & 11.57 & 13.35 & 35.05 & 31.10 & 30.24 \\
100 & 88.66 & 97.51 & 104.38 & 9.25 & 17.20 & 22.83 & 4.80 & 6.91 & 8.27 & 34.31 & 28.62 & 26.62 \\
150 & 138.84 & 145.65 & 155.11 & 7.50 & 13.52 & 18.61 & 3.93 & 5.16 & 6.54 & 34.39 & 27.47 & 26.03 \\
\hline \multicolumn{13}{l}{Adaptive LASSO (AIC)} \\ \hline 
50 & 47.50 & 18.82 & 15.87 & 97.47 & 73.57 & 76.26 & 94.75 & 34.08 & 27.33 & 49.29 & 31.61 & 26.35 \\
100 & 39.95 & 96.88 & 48.42 & 57.66 & 97.02 & 79.89 & 39.12 & 96.88 & 46.85 & 40.46 & 49.96 & 36.94 \\
150 & 42.79 & 68.41 & 146.27 & 49.03 & 58.35 & 96.56 & 27.93 & 45.18 & 97.55 & 36.36 & 43.65 & 50.25 \\
\hline \multicolumn{13}{l}{Adaptive LASSO (AICC)} \\ \hline 
50 & 26.15 & 11.29 & 12.04 & 69.39 & 66.23 & 73.02 & 50.62 & 18.26 & 19.24 & 41.82 & 21.63 & 20.85 \\
100 & 7.63 & 36.35 & 16.64 & 35.19 & 61.93 & 69.12 & 6.25 & 35.07 & 14.03 & 15.17 & 35.69 & 16.90 \\
150 & 10.44 & 10.29 & 41.89 & 32.54 & 41.31 & 59.00 & 6.13 & 5.71 & 26.90 & 15.91 & 12.20 & 30.93 \\
\hline \multicolumn{13}{l}{Adaptive LASSO (BIC)} \\ \hline 
50 & 37.89 & 5.97 & 6.26 & 82.13 & 50.64 & 59.51 & 75.15 & 8.12 & 7.87 & 47.72 & 13.86 & 11.70 \\
100 & 15.98 & 45.92 & 6.48 & 39.02 & 63.81 & 55.45 & 14.85 & 45.02 & 4.04 & 27.51 & 40.99 & 6.82 \\
150 & 29.00 & 5.19 & 30.76 & 42.05 & 37.06 & 52.93 & 18.88 & 2.34 & 19.43 & 30.99 & 6.00 & 26.38 \\
\hline \multicolumn{13}{l}{Oracle OLS} \\ \hline 
50 & 4.90 & 4.90 & 4.90 & 100.00 & 100.00 & 100.00 & 0.00 & 0.00 & 0.00 & 0.00 & 0.00 & 0.00 \\
100 & 4.95 & 4.95 & 4.95 & 100.00 & 100.00 & 100.00 & 0.00 & 0.00 & 0.00 & 0.00 & 0.00 & 0.00 \\
150 & 4.97 & 4.97 & 4.97 & 100.00 & 100.00 & 100.00 & 0.00 & 0.00 & 0.00 & 0.00 & 0.00 & 0.00 \\
  \hline \multicolumn{13}{l}{Number of Factors: 1} \\ \hline \hline \multicolumn{13}{l}{rigorous Lasso (HAC robust)} \\ \hline 
50 & 56.47 & 59.64 & 62.30 & 13.90 & 21.08 & 25.80 & 15.02 & 19.72 & 23.01 & 52.44 & 48.64 & 47.36 \\
100 & 111.73 & 116.91 & 120.90 & 9.04 & 14.13 & 18.32 & 8.97 & 12.05 & 14.29 & 50.17 & 46.36 & 44.06 \\
150 & 167.74 & 174.48 & 178.21 & 7.02 & 11.29 & 14.55 & 6.47 & 9.00 & 10.55 & 48.30 & 44.66 & 42.20 \\
\hline \multicolumn{13}{l}{Adaptive LASSO (AIC)} \\ \hline 
50 & 47.28 & 18.81 & 16.12 & 98.25 & 72.03 & 73.34 & 94.20 & 34.21 & 28.17 & 48.95 & 32.16 & 27.74 \\
100 & 24.42 & 96.12 & 48.11 & 63.02 & 98.48 & 78.07 & 22.54 & 96.00 & 46.61 & 26.06 & 49.36 & 37.36 \\
150 & 28.28 & 29.57 & 144.64 & 56.98 & 65.58 & 98.33 & 17.97 & 18.18 & 96.36 & 24.07 & 21.50 & 49.49 \\
\hline \multicolumn{13}{l}{Adaptive LASSO (AICC)} \\ \hline 
50 & 31.70 & 11.57 & 12.51 & 88.35 & 65.54 & 70.56 & 60.93 & 18.94 & 20.52 & 40.04 & 22.47 & 22.55 \\
100 & 7.48 & 32.58 & 17.34 & 63.69 & 80.47 & 66.82 & 4.68 & 30.19 & 14.87 & 7.27 & 26.00 & 18.23 \\
150 & 7.99 & 10.87 & 27.38 & 60.75 & 69.39 & 76.41 & 3.49 & 5.18 & 16.32 & 5.86 & 7.26 & 16.91 \\
\hline \multicolumn{13}{l}{Adaptive LASSO (BIC)} \\ \hline 
50 & 33.99 & 7.38 & 7.56 & 89.10 & 59.43 & 64.52 & 65.89 & 10.33 & 10.24 & 42.04 & 14.91 & 13.77 \\
100 & 7.90 & 20.62 & 8.38 & 64.87 & 80.84 & 61.95 & 5.06 & 17.62 & 5.71 & 7.79 & 17.24 & 8.52 \\
150 & 9.77 & 6.38 & 9.58 & 62.29 & 75.84 & 80.37 & 4.85 & 1.87 & 3.92 & 7.88 & 2.59 & 4.90 \\
\hline \multicolumn{13}{l}{Oracle OLS} \\ \hline 
50 & 4.90 & 4.90 & 4.90 & 100.00 & 100.00 & 100.00 & 0.00 & 0.00 & 0.00 & 0.00 & 0.00 & 0.00 \\
100 & 4.95 & 4.95 & 4.95 & 100.00 & 100.00 & 100.00 & 0.00 & 0.00 & 0.00 & 0.00 & 0.00 & 0.00 \\
150 & 4.97 & 4.97 & 4.97 & 100.00 & 100.00 & 100.00 & 0.00 & 0.00 & 0.00 & 0.00 & 0.00 & 0.00 \\
  \hline \multicolumn{13}{l}{Number of Factors: 5} \\ \hline \hline \multicolumn{13}{l}{rigorous Lasso (HAC robust)} \\ \hline 
50 & 68.33 & 71.25 & 72.83 & 30.42 & 38.10 & 42.27 & 19.56 & 24.92 & 28.34 & 39.28 & 39.62 & 40.18 \\
100 & 134.52 & 139.41 & 142.65 & 23.29 & 30.23 & 34.33 & 11.78 & 15.30 & 17.64 & 33.70 & 33.68 & 33.99 \\
150 & 198.68 & 206.49 & 211.56 & 19.89 & 26.69 & 31.06 & 8.62 & 11.32 & 13.18 & 30.32 & 29.85 & 29.85 \\
\hline \multicolumn{13}{l}{Adaptive LASSO (AIC)} \\ \hline 
50 & 7.58 & 6.88 & 6.32 & 92.77 & 65.51 & 67.93 & 7.48 & 8.64 & 7.17 & 7.23 & 11.26 & 9.26 \\
100 & 4.03 & 7.71 & 8.60 & 80.84 & 96.02 & 68.01 & 0.20 & 3.30 & 5.63 & 0.24 & 3.27 & 7.34 \\
150 & 4.02 & 4.14 & 7.67 & 80.43 & 82.76 & 97.02 & 0.09 & 0.10 & 2.05 & 0.12 & 0.12 & 2.06 \\
\hline \multicolumn{13}{l}{Adaptive LASSO (AICC)} \\ \hline 
50 & 5.62 & 6.28 & 6.08 & 92.44 & 64.50 & 67.54 & 3.22 & 7.42 & 6.69 & 3.36 & 10.11 & 8.80 \\
100 & 4.01 & 6.32 & 7.54 & 80.51 & 96.06 & 67.09 & 0.19 & 1.84 & 4.57 & 0.24 & 1.88 & 6.25 \\
150 & 4.00 & 4.14 & 6.58 & 80.08 & 82.66 & 97.10 & 0.09 & 0.10 & 1.30 & 0.11 & 0.12 & 1.32 \\
\hline \multicolumn{13}{l}{Adaptive LASSO (BIC)} \\ \hline 
50 & 5.52 & 5.33 & 5.27 & 91.73 & 62.42 & 65.90 & 3.07 & 5.54 & 5.06 & 3.23 & 8.10 & 7.04 \\
100 & 3.96 & 5.62 & 6.04 & 79.58 & 95.93 & 65.36 & 0.19 & 1.12 & 3.08 & 0.23 & 1.15 & 4.47 \\
150 & 3.95 & 4.09 & 5.69 & 79.11 & 81.74 & 97.22 & 0.09 & 0.09 & 0.68 & 0.11 & 0.12 & 0.69 \\
\hline \multicolumn{13}{l}{Oracle OLS} \\ \hline 
50 & 4.90 & 4.90 & 4.90 & 100.00 & 100.00 & 100.00 & 0.00 & 0.00 & 0.00 & 0.00 & 0.00 & 0.00 \\
100 & 4.95 & 4.95 & 4.95 & 100.00 & 100.00 & 100.00 & 0.00 & 0.00 & 0.00 & 0.00 & 0.00 & 0.00 \\
150 & 4.97 & 4.97 & 4.97 & 100.00 & 100.00 & 100.00 & 0.00 & 0.00 & 0.00 & 0.00 & 0.00 & 0.00 \\
\hline \hline \end{tabular} \caption{Monte Carlo Results for NSS and Specification 1. The number of dominant Units is set to 5. The parametrisation is summarised in Table \ref{tab:MC_specs}. The estimated equations are \eqref{eq:MC_dom} and \eqref{eq:MC_nondom}. The initial stage for the adaptive LASSO is a univariate OLS regression if \(N<T\) and multivariate if \(N>T\).}
\label{tab:tab_speclasso1_5_5} 
\end{table}

\clearpage
 \begin{table}[h]     \scriptsize     \centering     \begin{tabular}{@{\extracolsep{4pt}} c c c c c c c c c c c c c}     \hline \hline  & \multicolumn{3}{c}{\(\hat{N_d}\)} & \multicolumn{3}{c}{TPR} & \multicolumn{3}{c}{FPR} & \multicolumn{3}{c}{FDR} \\ \cline{2-4} \cline{5-7} \cline{8-10} \cline{11-13} N/T & 50 &100 &150 & 50 &100 &150 & 50 &100 &150 & 50 &100 &150 \\ \hline \hline \multicolumn{13}{l}{Number of Factors: 0} \\ \hline \hline \multicolumn{13}{l}{rigorous Lasso (HAC robust)} \\ \hline 
50 & 3.28 & 4.58 & 4.93 & 63.80 & 91.40 & 98.18 & 0.20 & 0.02 & 0.04 & 0.21 & 0.03 & 0.04 \\
100 & 3.32 & 4.65 & 4.99 & 62.80 & 92.80 & 99.80 & 0.19 & 0.01 & 0.00 & 0.35 & 0.01 & 0.00 \\
150 & 3.38 & 4.73 & 4.95 & 66.60 & 94.40 & 99.00 & 0.03 & 0.01 & 0.00 & 0.04 & 0.01 & 0.00 \\
\hline \multicolumn{13}{l}{Adaptive Lasso (AIC)} \\ \hline 
50 & 2.50 & 4.99 & 5.00 & 42.24 & 99.68 & 100.00 & 0.87 & 0.01 & 0.00 & 1.59 & 0.01 & 0.00 \\
100 & 4.88 & 2.51 & 5.00 & 96.40 & 47.80 & 99.96 & 0.06 & 0.13 & 0.00 & 0.07 & 0.96 & 0.00 \\
150 & 4.60 & 5.00 & 2.76 & 89.40 & 99.92 & 54.20 & 0.09 & 0.00 & 0.04 & 0.16 & 0.00 & 0.24 \\
\hline \multicolumn{13}{l}{Adaptive Lasso (AICC)} \\ \hline 
50 & 2.67 & 5.00 & 5.00 & 40.88 & 100.00 & 100.00 & 1.39 & 0.00 & 0.00 & 2.43 & 0.00 & 0.00 \\
100 & 4.98 & 2.39 & 5.00 & 98.64 & 43.76 & 100.00 & 0.05 & 0.21 & 0.00 & 0.05 & 1.32 & 0.00 \\
150 & 4.75 & 5.00 & 2.54 & 94.80 & 100.00 & 48.88 & 0.01 & 0.00 & 0.07 & 0.01 & 0.00 & 0.54 \\
\hline \multicolumn{13}{l}{Adaptive Lasso (BIC)} \\ \hline 
50 & 2.58 & 5.00 & 5.00 & 42.68 & 100.00 & 100.00 & 0.98 & 0.00 & 0.00 & 1.66 & 0.00 & 0.00 \\
100 & 4.96 & 2.78 & 5.00 & 99.00 & 53.16 & 100.00 & 0.01 & 0.13 & 0.00 & 0.01 & 0.18 & 0.00 \\
150 & 4.74 & 5.00 & 4.24 & 92.68 & 100.00 & 84.52 & 0.07 & 0.00 & 0.01 & 0.15 & 0.00 & 0.01 \\
\hline \multicolumn{13}{l}{Brownlees and Mesters} \\ \hline 
50 & 4.96 & 5.00 & 5.00 & 99.04 & 100.00 & 100.00 & 0.02 & 0.00 & 0.00 & 0.02 & 0.00 & 0.00 \\
100 & . & 5.00 & 5.00 & . & 100.00 & 100.00 & . & 0.00 & 0.00 & . & 0.00 & 0.00 \\
150 & . & . & 5.00 & . & . & 100.00 & . & . & 0.00 & . & . & 0.00 \\
  \hline \multicolumn{13}{l}{Number of Factors: 1} \\ \hline \hline \multicolumn{13}{l}{rigorous Lasso (HAC robust)} \\ \hline 
50 & 2.78 & 4.06 & 4.33 & 54.63 & 80.80 & 86.60 & 0.11 & 0.04 & 0.00 & 0.16 & 0.04 & 0.00 \\
100 & 3.16 & 4.21 & 4.74 & 62.40 & 84.20 & 94.80 & 0.04 & 0.00 & 0.00 & 0.05 & 0.00 & 0.00 \\
150 & 3.24 & 4.49 & 4.71 & 64.20 & 89.60 & 94.20 & 0.02 & 0.01 & 0.00 & 0.02 & 0.01 & 0.00 \\
\hline \multicolumn{13}{l}{Adaptive Lasso (AIC)} \\ \hline 
50 & 2.53 & 4.96 & 5.00 & 41.12 & 99.20 & 100.00 & 1.06 & 0.01 & 0.00 & 2.58 & 0.01 & 0.00 \\
100 & 4.71 & 2.65 & 4.99 & 94.12 & 49.44 & 99.80 & 0.00 & 0.19 & 0.00 & 0.00 & 0.27 & 0.00 \\
150 & 4.72 & 4.91 & 3.17 & 94.32 & 98.12 & 62.80 & 0.01 & 0.00 & 0.02 & 0.01 & 0.00 & 0.02 \\
\hline \multicolumn{13}{l}{Adaptive Lasso (AICC)} \\ \hline 
50 & 2.64 & 4.99 & 5.00 & 41.20 & 99.76 & 100.00 & 1.30 & 0.00 & 0.00 & 3.04 & 0.00 & 0.00 \\
100 & 4.83 & 3.26 & 5.00 & 96.52 & 59.04 & 100.00 & 0.01 & 0.32 & 0.00 & 0.01 & 1.02 & 0.00 \\
150 & 4.80 & 4.95 & 3.77 & 96.04 & 99.00 & 74.88 & 0.00 & 0.00 & 0.02 & 0.00 & 0.00 & 0.04 \\
\hline \multicolumn{13}{l}{Adaptive Lasso (BIC)} \\ \hline 
50 & 2.69 & 5.00 & 5.00 & 47.24 & 99.92 & 100.00 & 0.73 & 0.00 & 0.00 & 1.69 & 0.00 & 0.00 \\
100 & 4.76 & 4.40 & 5.00 & 95.12 & 87.80 & 100.00 & 0.01 & 0.01 & 0.00 & 0.01 & 0.01 & 0.00 \\
150 & 4.73 & 4.89 & 4.88 & 94.52 & 97.84 & 97.60 & 0.00 & 0.00 & 0.00 & 0.00 & 0.00 & 0.00 \\
\hline \multicolumn{13}{l}{BM procedure} \\ \hline 
50 & 4.54 & 4.73 & 4.81 & 90.44 & 94.56 & 96.08 & 0.04 & 0.01 & 0.01 & 0.04 & 0.01 & 0.01 \\
100 & . & 4.40 & 4.51 & . & 87.48 & 89.84 & . & 0.02 & 0.01 & . & 0.02 & 0.01 \\
150 & . & . & 4.27 & . & . & 85.24 & . & . & 0.01 & . & . & 0.01 \\
  \hline \multicolumn{13}{l}{Number of Factors: 5} \\ \hline \hline \multicolumn{13}{l}{rigorous Lasso (HAC robust)} \\ \hline 
50 & 4.77 & 4.97 & 5.00 & 94.80 & 99.40 & 100.00 & 0.07 & 0.00 & 0.00 & 0.07 & 0.00 & 0.00 \\
100 & 4.90 & 4.98 & 5.00 & 97.00 & 99.60 & 100.00 & 0.05 & 0.00 & 0.00 & 0.05 & 0.00 & 0.00 \\
150 & 4.77 & 4.98 & 5.00 & 94.80 & 99.60 & 100.00 & 0.02 & 0.00 & 0.00 & 0.02 & 0.00 & 0.00 \\
\hline \multicolumn{13}{l}{Adaptive Lasso (AIC)} \\ \hline 
50 & 5.05 & 5.00 & 5.00 & 99.92 & 99.92 & 100.00 & 0.11 & 0.00 & 0.00 & 0.11 & 0.00 & 0.00 \\
100 & 3.96 & 5.15 & 5.00 & 78.92 & 100.00 & 100.00 & 0.01 & 0.16 & 0.00 & 0.01 & 0.15 & 0.00 \\
150 & 4.10 & 4.08 & 5.22 & 81.24 & 81.12 & 100.00 & 0.03 & 0.02 & 0.15 & 0.03 & 0.02 & 0.15 \\
\hline \multicolumn{13}{l}{Adaptive Lasso (AICC)} \\ \hline 
50 & 5.05 & 5.00 & 5.00 & 99.92 & 99.96 & 100.00 & 0.12 & 0.00 & 0.00 & 0.12 & 0.00 & 0.00 \\
100 & 3.92 & 5.17 & 5.00 & 78.24 & 100.00 & 100.00 & 0.01 & 0.17 & 0.00 & 0.01 & 0.17 & 0.00 \\
150 & 4.05 & 4.08 & 5.21 & 80.44 & 81.00 & 100.00 & 0.02 & 0.02 & 0.15 & 0.02 & 0.02 & 0.15 \\
\hline \multicolumn{13}{l}{Adaptive Lasso (BIC)} \\ \hline 
50 & 5.04 & 5.00 & 5.00 & 99.92 & 99.92 & 100.00 & 0.10 & 0.00 & 0.00 & 0.10 & 0.00 & 0.00 \\
100 & 3.81 & 5.11 & 5.00 & 76.04 & 100.00 & 100.00 & 0.01 & 0.11 & 0.00 & 0.01 & 0.11 & 0.00 \\
150 & 3.97 & 3.97 & 5.20 & 78.84 & 78.96 & 100.00 & 0.02 & 0.01 & 0.14 & 0.02 & 0.01 & 0.14 \\
\hline \multicolumn{13}{l}{BM procedure} \\ \hline 
50 & 4.75 & 4.83 & 4.70 & 95.00 & 96.20 & 93.88 & 0.00 & 0.04 & 0.02 & 0.00 & 0.03 & 0.02 \\
100 & . & 4.85 & 4.85 & . & 96.48 & 96.60 & . & 0.03 & 0.03 & . & 0.03 & 0.03 \\
150 & . & . & 4.78 & . & . & 94.80 & . & . & 0.03 & . & . & 0.03 \\
\hline \hline \end{tabular} \caption{Monte Carlo Results for estimation of Dominant Drivers, Specification 1. The number of dominant Units is 5. See notes of Table \ref{tab:tab_speclasso1_5_5}. \textit{BM procedure} is the procedure in \cite{Brownlees2021} directly applied to the inverse of the simulated sample covariance matrix.} \label{tab:tab_spec1_5_5} \end{table}

\clearpage
 \begin{table}[h]     \scriptsize     \centering     \begin{tabular}{@{\extracolsep{4pt}} c c c c c c c c c c c c c}     \hline \hline  & \multicolumn{3}{c}{\(\hat{N_d}\)} & \multicolumn{3}{c}{TPR} & \multicolumn{3}{c}{FPR} & \multicolumn{3}{c}{FDR} \\ \cline{2-4} \cline{5-7} \cline{8-10} \cline{11-13} N/T & 50 &100 &150 & 50 &100 &150 & 50 &100 &150 & 50 &100 &150 \\ \hline \hline \multicolumn{13}{l}{Number of Factors: 0} \\ \hline \hline \multicolumn{13}{l}{rigorous LASSO (HAC robust)} \\ \hline 
50 & 3.14 & 2.90 & 3.48 & 48.40 & 55.60 & 66.80 & 1.60 & 0.27 & 0.31 & 2.51 & 0.50 & 0.29 \\
100 & 2.20 & 3.06 & 2.82 & 41.20 & 57.20 & 55.60 & 0.15 & 0.21 & 0.04 & 0.29 & 0.22 & 0.07 \\
150 & 2.88 & 2.90 & 3.18 & 54.00 & 57.20 & 63.60 & 0.12 & 0.03 & 0.00 & 0.15 & 0.05 & 0.00 \\
\hline \multicolumn{13}{l}{Adaptive LASSO (AIC)} \\ \hline 
50 & 2.95 & 4.04 & 4.58 & 33.76 & 79.84 & 91.08 & 2.81 & 0.11 & 0.05 & 9.19 & 0.11 & 0.05 \\
100 & 4.44 & 2.55 & 4.32 & 88.00 & 34.80 & 85.20 & 0.04 & 0.85 & 0.06 & 0.06 & 5.65 & 0.06 \\
150 & 4.46 & 4.95 & 2.39 & 76.24 & 99.00 & 36.92 & 0.44 & 0.00 & 0.37 & 2.00 & 0.00 & 2.46 \\
\hline \multicolumn{13}{l}{Adaptive LASSO (AICC)} \\ \hline 
50 & 3.01 & 4.20 & 4.63 & 33.20 & 82.88 & 91.56 & 3.00 & 0.12 & 0.11 & 10.56 & 0.12 & 0.10 \\
100 & 4.24 & 2.91 & 4.76 & 83.60 & 34.04 & 94.68 & 0.07 & 1.28 & 0.03 & 0.28 & 8.30 & 0.03 \\
150 & 4.40 & 4.97 & 2.46 & 80.80 & 99.36 & 38.80 & 0.25 & 0.00 & 0.36 & 1.11 & 0.00 & 3.87 \\
\hline \multicolumn{13}{l}{Adaptive LASSO (BIC)} \\ \hline 
50 & 3.03 & 4.36 & 4.66 & 33.72 & 83.28 & 89.44 & 2.98 & 0.43 & 0.43 & 9.08 & 0.40 & 0.40 \\
100 & 4.35 & 2.48 & 4.83 & 86.16 & 35.44 & 95.76 & 0.05 & 0.74 & 0.05 & 0.09 & 5.35 & 0.05 \\
150 & 4.47 & 4.83 & 2.39 & 79.72 & 96.20 & 43.12 & 0.33 & 0.01 & 0.16 & 1.82 & 0.01 & 1.46 \\
\hline \multicolumn{13}{l}{BM procedure} \\ \hline 
50 & 4.83 & 4.99 & 5.00 & 95.72 & 99.84 & 100.00 & 0.09 & 0.00 & 0.00 & 0.09 & 0.00 & 0.00 \\
100 & . & 4.97 & 5.00 & . & 99.40 & 100.00 & . & 0.00 & 0.00 & . & 0.00 & 0.00 \\
150 & . & . & 5.00 & . & . & 99.96 & . & . & 0.00 & . & . & 0.00 \\
  \hline \multicolumn{13}{l}{Number of Factors: 1} \\ \hline \hline \multicolumn{13}{l}{rigorous LASSO (HAC robust)} \\ \hline 
50 & 4.56 & 2.82 & 4.76 & 45.20 & 42.00 & 53.20 & 5.11 & 1.60 & 4.67 & 9.96 & 2.13 & 5.79 \\
100 & 3.94 & 2.90 & 2.60 & 36.00 & 46.00 & 51.20 & 2.25 & 0.63 & 0.04 & 6.41 & 2.92 & 0.05 \\
150 & 2.14 & 2.78 & 3.36 & 32.80 & 46.40 & 64.00 & 0.34 & 0.32 & 0.11 & 2.71 & 0.50 & 0.13 \\
\hline \multicolumn{13}{l}{Adaptive LASSO (AIC)} \\ \hline 
50 & 2.63 & 3.88 & 4.54 & 31.80 & 75.64 & 89.76 & 2.32 & 0.22 & 0.12 & 8.38 & 0.23 & 0.11 \\
100 & 3.95 & 2.53 & 3.97 & 78.04 & 33.00 & 78.36 & 0.05 & 0.93 & 0.05 & 0.05 & 6.76 & 0.05 \\
150 & 4.24 & 4.64 & 2.35 & 72.48 & 92.64 & 36.12 & 0.42 & 0.01 & 0.37 & 2.34 & 0.01 & 3.49 \\
\hline \multicolumn{13}{l}{Adaptive LASSO (AICC)} \\ \hline 
50 & 3.15 & 4.15 & 4.49 & 32.96 & 80.96 & 88.96 & 3.33 & 0.23 & 0.09 & 11.06 & 0.24 & 0.09 \\
100 & 4.51 & 2.56 & 4.62 & 87.96 & 32.64 & 91.96 & 0.12 & 0.98 & 0.03 & 0.12 & 7.07 & 0.03 \\
150 & 4.48 & 4.96 & 2.60 & 85.36 & 99.20 & 39.80 & 0.15 & 0.00 & 0.42 & 0.58 & 0.00 & 4.91 \\
\hline \multicolumn{13}{l}{Adaptive LASSO (BIC)} \\ \hline 
50 & 2.76 & 3.67 & 4.00 & 32.80 & 71.36 & 78.52 & 2.48 & 0.23 & 0.16 & 8.63 & 0.22 & 0.16 \\
100 & 4.19 & 2.29 & 4.27 & 82.88 & 36.68 & 83.60 & 0.05 & 0.48 & 0.09 & 0.05 & 3.23 & 0.09 \\
150 & 4.35 & 4.09 & 3.17 & 79.44 & 81.28 & 61.12 & 0.26 & 0.02 & 0.08 & 1.62 & 0.02 & 0.29 \\
\hline \multicolumn{13}{l}{BM procedure} \\ \hline 
50 & 2.69 & 2.89 & 3.08 & 50.68 & 57.24 & 61.60 & 0.36 & 0.07 & 0.01 & 1.07 & 0.28 & 0.01 \\
100 & . & 2.49 & 2.56 & . & 49.24 & 50.68 & . & 0.03 & 0.03 & . & 0.86 & 0.47 \\
150 & . & . & 2.46 & . & . & 48.00 & . & . & 0.04 & . & . & 0.84 \\
  \hline \multicolumn{13}{l}{Number of Factors: 5} \\ \hline \hline \multicolumn{13}{l}{rigorous LASSO (HAC robust)} \\ \hline 
50 & 3.38 & 4.28 & 4.54 & 66.00 & 85.20 & 90.40 & 0.18 & 0.04 & 0.04 & 0.20 & 0.04 & 0.04 \\
100 & 3.02 & 4.62 & 4.82 & 59.60 & 91.60 & 96.00 & 0.04 & 0.04 & 0.02 & 0.05 & 0.04 & 0.02 \\
150 & 3.60 & 4.38 & 4.88 & 70.00 & 87.20 & 97.60 & 0.07 & 0.01 & 0.00 & 0.07 & 0.01 & 0.00 \\
\hline \multicolumn{13}{l}{Adaptive LASSO (AIC)} \\ \hline 
50 & 2.92 & 3.44 & 3.82 & 40.56 & 67.44 & 75.64 & 1.98 & 0.14 & 0.08 & 5.02 & 0.15 & 0.08 \\
100 & 3.22 & 2.35 & 3.58 & 26.76 & 40.44 & 70.76 & 1.98 & 0.34 & 0.04 & 10.46 & 1.99 & 0.04 \\
150 & 4.52 & 3.04 & 2.47 & 22.48 & 30.68 & 46.00 & 2.34 & 1.04 & 0.12 & 16.97 & 5.43 & 0.16 \\
\hline \multicolumn{13}{l}{Adaptive LASSO (AICC)} \\ \hline 
50 & 2.44 & 3.54 & 3.82 & 36.48 & 68.44 & 75.92 & 1.38 & 0.27 & 0.06 & 1.91 & 0.27 & 0.06 \\
100 & 3.87 & 1.96 & 4.21 & 26.80 & 34.52 & 83.12 & 2.66 & 0.25 & 0.06 & 13.65 & 1.18 & 0.06 \\
150 & 4.46 & 3.05 & 2.12 & 22.12 & 28.92 & 33.12 & 2.31 & 1.11 & 0.32 & 18.93 & 6.88 & 1.20 \\
\hline \multicolumn{13}{l}{Adaptive LASSO (BIC)} \\ \hline 
50 & 2.47 & 3.90 & 3.99 & 35.72 & 76.04 & 78.80 & 1.52 & 0.21 & 0.11 & 2.40 & 0.21 & 0.10 \\
100 & 3.64 & 2.82 & 4.48 & 26.12 & 31.96 & 88.44 & 2.46 & 1.28 & 0.06 & 12.48 & 2.82 & 0.06 \\
150 & 4.61 & 3.91 & 2.94 & 22.20 & 26.40 & 30.72 & 2.42 & 1.78 & 0.97 & 17.68 & 9.14 & 3.59 \\
\hline \multicolumn{13}{l}{BM procedure} \\ \hline 
50 & 2.63 & 2.49 & 2.53 & 8.08 & 6.96 & 6.48 & 4.95 & 4.75 & 4.91 & 76.20 & 76.80 & 79.51 \\
100 & . & 2.32 & 2.26 & . & 2.52 & 3.12 & . & 2.31 & 2.22 & . & 90.61 & 87.57 \\
150 & . & . & 2.42 & . & . & 1.76 & . & . & 1.61 & . & . & 92.50 \\
\hline \hline \end{tabular} \caption{Monte Carlo Results for estimation of Dominant Drivers, Specification 2. The number of dominant Units is 5. See notes of Table \ref{tab:tab_speclasso1_5_5} and Table \ref{tab:tab_spec1_5_5}.} \label{tab:tab_spec2_5_5} \end{table}

\clearpage
 \begin{table}[h]     \scriptsize     \centering     \begin{tabular}{@{\extracolsep{4pt}} c c c c c c c c c c c c c}     \hline \hline  & \multicolumn{3}{c}{\(\hat{N_d}\)} & \multicolumn{3}{c}{TPR} & \multicolumn{3}{c}{FPR} & \multicolumn{3}{c}{FDR} \\ \cline{2-4} \cline{5-7} \cline{8-10} \cline{11-13} N/T & 50 &100 &150 & 50 &100 &150 & 50 &100 &150 & 50 &100 &150 \\ \hline \hline \multicolumn{13}{l}{Number of Factors: 0} \\ \hline \hline \multicolumn{13}{l}{rigorous LASSO (HAC robust)} \\ \hline 
50 & 3.54 & 4.52 & 5.00 & 60.40 & 90.00 & 99.60 & 1.16 & 0.04 & 0.04 & 1.36 & 0.04 & 0.04 \\
100 & 2.94 & 4.36 & 4.96 & 57.20 & 87.20 & 99.20 & 0.08 & 0.00 & 0.00 & 0.12 & 0.00 & 0.00 \\
150 & 3.54 & 4.84 & 4.98 & 64.40 & 96.80 & 99.60 & 0.22 & 0.00 & 0.00 & 0.23 & 0.00 & 0.00 \\
\hline \multicolumn{13}{l}{Adaptive LASSO (AIC)} \\ \hline 
50 & 2.72 & 4.99 & 5.00 & 50.40 & 99.76 & 100.00 & 0.45 & 0.01 & 0.00 & 0.48 & 0.01 & 0.00 \\
100 & 4.99 & 3.26 & 5.00 & 99.88 & 64.68 & 99.96 & 0.00 & 0.03 & 0.00 & 0.00 & 0.04 & 0.00 \\
150 & 4.96 & 5.00 & 4.09 & 99.16 & 100.00 & 81.44 & 0.00 & 0.00 & 0.01 & 0.01 & 0.00 & 0.01 \\
\hline \multicolumn{13}{l}{Adaptive LASSO (AICC)} \\ \hline 
50 & 2.81 & 5.00 & 5.00 & 51.60 & 99.96 & 100.00 & 0.51 & 0.00 & 0.00 & 1.08 & 0.00 & 0.00 \\
100 & 5.00 & 3.89 & 5.00 & 99.84 & 77.36 & 100.00 & 0.00 & 0.02 & 0.00 & 0.00 & 0.03 & 0.00 \\
150 & 4.94 & 5.00 & 4.85 & 98.80 & 100.00 & 96.80 & 0.00 & 0.00 & 0.01 & 0.00 & 0.00 & 0.01 \\
\hline \multicolumn{13}{l}{Adaptive LASSO (BIC)} \\ \hline 
50 & 2.79 & 5.00 & 5.00 & 51.88 & 99.88 & 100.00 & 0.44 & 0.00 & 0.00 & 0.44 & 0.00 & 0.00 \\
100 & 5.00 & 4.16 & 5.00 & 99.96 & 82.04 & 100.00 & 0.00 & 0.06 & 0.00 & 0.00 & 0.06 & 0.00 \\
150 & 5.03 & 5.00 & 4.99 & 99.44 & 100.00 & 99.80 & 0.04 & 0.00 & 0.00 & 0.03 & 0.00 & 0.00 \\
\hline \multicolumn{13}{l}{BM procedure} \\ \hline 
50 & 4.98 & 5.00 & 5.00 & 99.60 & 100.00 & 100.00 & 0.00 & 0.00 & 0.00 & 0.00 & 0.00 & 0.00 \\
100 & . & 5.00 & 5.00 & . & 100.00 & 100.00 & . & 0.00 & 0.00 & . & 0.00 & 0.00 \\
150 & . & . & 5.00 & . & . & 100.00 & . & . & 0.00 & . & . & 0.00 \\
  \hline \multicolumn{13}{l}{Number of Factors: 1} \\ \hline \hline \multicolumn{13}{l}{rigorous LASSO (HAC robust)} \\ \hline 
50 & 2.96 & 3.98 & 4.38 & 54.80 & 79.60 & 87.60 & 0.49 & 0.00 & 0.00 & 0.61 & 0.00 & 0.00 \\
100 & 3.26 & 4.38 & 4.54 & 65.20 & 87.20 & 90.80 & 0.00 & 0.02 & 0.00 & 0.00 & 0.03 & 0.00 \\
150 & 3.52 & 4.30 & 4.78 & 68.80 & 86.00 & 95.60 & 0.06 & 0.00 & 0.00 & 0.07 & 0.00 & 0.00 \\
\hline \multicolumn{13}{l}{Adaptive LASSO (AIC)} \\ \hline 
50 & 2.44 & 4.93 & 5.00 & 46.24 & 98.56 & 99.96 & 0.29 & 0.01 & 0.00 & 0.43 & 0.01 & 0.00 \\
100 & 4.93 & 3.25 & 4.99 & 98.40 & 64.64 & 99.80 & 0.01 & 0.02 & 0.00 & 0.01 & 0.03 & 0.00 \\
150 & 4.85 & 4.98 & 4.18 & 97.00 & 99.68 & 83.28 & 0.00 & 0.00 & 0.01 & 0.00 & 0.00 & 0.01 \\
\hline \multicolumn{13}{l}{Adaptive LASSO (AICC)} \\ \hline 
50 & 2.70 & 5.01 & 5.00 & 49.64 & 100.00 & 99.96 & 0.48 & 0.02 & 0.00 & 0.88 & 0.02 & 0.00 \\
100 & 4.94 & 3.90 & 5.00 & 98.84 & 77.56 & 99.96 & 0.00 & 0.02 & 0.00 & 0.00 & 0.04 & 0.00 \\
150 & 4.92 & 4.98 & 4.69 & 98.44 & 99.64 & 93.72 & 0.00 & 0.00 & 0.00 & 0.00 & 0.00 & 0.20 \\
\hline \multicolumn{13}{l}{Adaptive LASSO (BIC)} \\ \hline 
50 & 3.00 & 4.97 & 5.00 & 57.36 & 99.40 & 100.00 & 0.28 & 0.00 & 0.00 & 0.37 & 0.00 & 0.00 \\
100 & 4.90 & 4.71 & 5.00 & 97.96 & 94.20 & 100.00 & 0.00 & 0.00 & 0.00 & 0.00 & 0.00 & 0.00 \\
150 & 4.88 & 4.95 & 4.97 & 97.56 & 99.08 & 99.40 & 0.00 & 0.00 & 0.00 & 0.00 & 0.00 & 0.00 \\
\hline \multicolumn{13}{l}{BM procedure} \\ \hline 
50 & 4.39 & 4.46 & 4.63 & 87.68 & 89.24 & 92.60 & 0.02 & 0.00 & 0.00 & 0.02 & 0.00 & 0.00 \\
100 & . & 4.13 & 4.36 & . & 82.28 & 87.12 & . & 0.01 & 0.00 & . & 0.01 & 0.00 \\
150 & . & . & 4.08 & . & . & 81.48 & . & . & 0.00 & . & . & 0.00 \\
  \hline \multicolumn{13}{l}{Number of Factors: 5} \\ \hline \hline \multicolumn{13}{l}{rigorous LASSO (HAC robust)} \\ \hline 
50 & 5.04 & 4.98 & 5.00 & 98.40 & 99.60 & 100.00 & 0.27 & 0.00 & 0.00 & 0.27 & 0.00 & 0.00 \\
100 & 4.72 & 5.00 & 5.00 & 94.00 & 100.00 & 100.00 & 0.02 & 0.00 & 0.00 & 0.02 & 0.00 & 0.00 \\
150 & 4.70 & 4.96 & 5.00 & 93.60 & 99.20 & 100.00 & 0.01 & 0.00 & 0.00 & 0.01 & 0.00 & 0.00 \\
\hline \multicolumn{13}{l}{Adaptive LASSO (AIC)} \\ \hline 
50 & 4.97 & 4.98 & 4.99 & 98.80 & 99.56 & 99.88 & 0.07 & 0.00 & 0.00 & 0.07 & 0.00 & 0.00 \\
100 & 3.25 & 5.03 & 5.00 & 64.88 & 99.24 & 100.00 & 0.01 & 0.08 & 0.00 & 0.01 & 0.07 & 0.00 \\
150 & 3.33 & 3.28 & 5.09 & 66.16 & 65.36 & 99.96 & 0.01 & 0.01 & 0.06 & 0.01 & 0.01 & 0.06 \\
\hline \multicolumn{13}{l}{Adaptive LASSO (AICC)} \\ \hline 
50 & 4.99 & 4.98 & 4.99 & 98.96 & 99.64 & 99.88 & 0.09 & 0.00 & 0.00 & 0.09 & 0.00 & 0.00 \\
100 & 3.24 & 5.05 & 5.00 & 64.60 & 99.24 & 100.00 & 0.01 & 0.09 & 0.00 & 0.01 & 0.09 & 0.00 \\
150 & 3.32 & 3.27 & 5.09 & 66.08 & 65.12 & 99.96 & 0.01 & 0.01 & 0.06 & 0.01 & 0.01 & 0.06 \\
\hline \multicolumn{13}{l}{Adaptive LASSO (BIC)} \\ \hline 
50 & 5.00 & 4.98 & 4.99 & 99.12 & 99.64 & 99.80 & 0.09 & 0.00 & 0.00 & 0.09 & 0.00 & 0.00 \\
100 & 3.12 & 5.04 & 5.00 & 62.36 & 99.28 & 100.00 & 0.01 & 0.08 & 0.00 & 0.01 & 0.07 & 0.00 \\
150 & 3.22 & 3.18 & 5.10 & 64.04 & 63.44 & 99.96 & 0.01 & 0.01 & 0.07 & 0.01 & 0.01 & 0.07 \\
\hline \multicolumn{13}{l}{BM procedure} \\ \hline 
50 & 4.81 & 4.78 & 4.83 & 96.04 & 95.52 & 96.36 & 0.02 & 0.01 & 0.03 & 0.02 & 0.01 & 0.03 \\
100 & . & 4.78 & 4.86 & . & 95.36 & 96.56 & . & 0.02 & 0.04 & . & 0.02 & 0.04 \\
150 & . & . & 4.78 & . & . & 94.80 & . & . & 0.03 & . & . & 0.03 \\
\hline \hline \end{tabular} \caption{Monte Carlo Results for estimation of Dominant Drivers, Specification 3. The number of dominant Units is 5. See notes of Table \ref{tab:tab_speclasso1_5_5} and Table \ref{tab:tab_spec1_5_5}.} \label{tab:tab_spec3_5_5} \end{table}

\clearpage
 \begin{table}[h]     \scriptsize     \centering     \begin{tabular}{@{\extracolsep{4pt}} c c c c c c c c c c c c c}     \hline \hline  & \multicolumn{3}{c}{\(\hat{N_d}\)} & \multicolumn{3}{c}{TPR} & \multicolumn{3}{c}{FPR} & \multicolumn{3}{c}{FDR} \\ \cline{2-4} \cline{5-7} \cline{8-10} \cline{11-13} N/T & 50 &100 &150 & 50 &100 &150 & 50 &100 &150 & 50 &100 &150 \\ \hline \hline \multicolumn{13}{l}{Number of Factors: 0} \\ \hline \hline \multicolumn{13}{l}{rigorous Lasso (HAC robust)} \\ \hline 
50 & 2.18 & 4.06 & 4.58 & 36.80 & 81.20 & 91.60 & 0.76 & 0.00 & 0.00 & 4.00 & 0.00 & 0.00 \\
100 & 2.66 & 2.58 & 2.18 & 11.20 & 21.20 & 38.00 & 2.21 & 1.60 & 0.29 & 60.57 & 35.22 & 6.49 \\
150 & 5.36 & 2.94 & 1.94 & 3.20 & 8.80 & 16.00 & 3.59 & 1.72 & 0.79 & 91.24 & 67.20 & 43.10 \\
\hline \multicolumn{13}{l}{Adaptive Lasso (AIC)} \\ \hline 
50 & 3.56 & 4.99 & 5.00 & 70.40 & 99.88 & 100.00 & 0.10 & 0.00 & 0.00 & 0.10 & 0.00 & 0.00 \\
100 & 5.00 & 4.27 & 5.00 & 99.96 & 85.32 & 100.00 & 0.00 & 0.00 & 0.00 & 0.00 & 0.00 & 0.00 \\
150 & 4.99 & 5.00 & 4.84 & 99.88 & 100.00 & 96.68 & 0.00 & 0.00 & 0.00 & 0.00 & 0.00 & 0.00 \\
\hline \multicolumn{13}{l}{Adaptive Lasso (AICC)} \\ \hline 
50 & 3.29 & 5.00 & 5.00 & 65.24 & 99.92 & 100.00 & 0.06 & 0.00 & 0.00 & 0.07 & 0.00 & 0.00 \\
100 & 5.00 & 4.55 & 5.00 & 100.00 & 90.76 & 100.00 & 0.00 & 0.01 & 0.00 & 0.00 & 0.01 & 0.00 \\
150 & 5.00 & 5.00 & 4.96 & 100.00 & 100.00 & 99.28 & 0.00 & 0.00 & 0.00 & 0.00 & 0.00 & 0.00 \\
\hline \multicolumn{13}{l}{Adaptive Lasso (BIC)} \\ \hline 
50 & 4.17 & 5.00 & 4.99 & 82.84 & 99.96 & 99.72 & 0.06 & 0.00 & 0.00 & 0.06 & 0.00 & 0.00 \\
100 & 5.00 & 5.00 & 5.00 & 100.00 & 100.00 & 100.00 & 0.00 & 0.00 & 0.00 & 0.00 & 0.00 & 0.00 \\
150 & 5.00 & 5.00 & 5.00 & 100.00 & 100.00 & 100.00 & 0.00 & 0.00 & 0.00 & 0.00 & 0.00 & 0.00 \\
\hline \multicolumn{13}{l}{Brownlees and Meesters} \\ \hline 
50 & 5.05 & 5.02 & 5.03 & 100.00 & 100.00 & 100.00 & 0.11 & 0.04 & 0.06 & 0.10 & 0.04 & 0.06 \\
100 & . & 5.04 & 5.05 & . & 100.00 & 100.00 & . & 0.04 & 0.05 & . & 0.04 & 0.05 \\
150 & . & . & 5.09 & . & . & 100.00 & . & . & 0.06 & . & . & 0.06 \\
  \hline \multicolumn{13}{l}{Number of Factors: 1} \\ \hline \hline \multicolumn{13}{l}{rigorous Lasso (HAC robust)} \\ \hline 
50 & 2.48 & 3.90 & 4.04 & 49.60 & 78.00 & 80.80 & 0.00 & 0.00 & 0.00 & 0.00 & 0.00 & 0.00 \\
100 & 1.68 & 2.52 & 2.84 & 33.60 & 50.40 & 56.80 & 0.00 & 0.00 & 0.00 & 0.00 & 0.00 & 0.00 \\
150 & 1.50 & 1.70 & 2.22 & 29.60 & 34.00 & 44.40 & 0.01 & 0.00 & 0.00 & 0.03 & 0.00 & 0.00 \\
\hline \multicolumn{13}{l}{Adaptive Lasso (AIC)} \\ \hline 
50 & 3.71 & 4.99 & 5.00 & 73.44 & 99.80 & 100.00 & 0.09 & 0.00 & 0.00 & 0.12 & 0.00 & 0.00 \\
100 & 4.92 & 4.54 & 5.00 & 98.24 & 90.80 & 99.96 & 0.00 & 0.00 & 0.00 & 0.00 & 0.00 & 0.00 \\
150 & 4.94 & 4.97 & 4.95 & 98.68 & 99.40 & 98.92 & 0.01 & 0.00 & 0.00 & 0.01 & 0.00 & 0.00 \\
\hline \multicolumn{13}{l}{Adaptive Lasso (AICC)} \\ \hline 
50 & 3.68 & 4.99 & 5.00 & 72.40 & 99.80 & 99.96 & 0.14 & 0.00 & 0.00 & 0.17 & 0.00 & 0.00 \\
100 & 4.81 & 4.79 & 5.00 & 96.04 & 95.72 & 99.96 & 0.01 & 0.00 & 0.00 & 0.01 & 0.00 & 0.00 \\
150 & 4.85 & 4.93 & 4.98 & 96.60 & 98.56 & 99.64 & 0.01 & 0.00 & 0.00 & 0.01 & 0.00 & 0.00 \\
\hline \multicolumn{13}{l}{Adaptive Lasso (BIC)} \\ \hline 
50 & 4.33 & 4.98 & 4.99 & 86.52 & 99.68 & 99.88 & 0.01 & 0.00 & 0.00 & 0.01 & 0.00 & 0.00 \\
100 & 4.73 & 4.83 & 5.00 & 94.48 & 96.48 & 100.00 & 0.01 & 0.01 & 0.00 & 0.01 & 0.01 & 0.00 \\
150 & 4.77 & 4.84 & 4.87 & 95.16 & 96.40 & 96.92 & 0.01 & 0.01 & 0.01 & 0.01 & 0.01 & 0.01 \\
\hline \multicolumn{13}{l}{Brownlees and Meesters} \\ \hline 
50 & 5.01 & 4.99 & 5.01 & 98.84 & 99.12 & 99.20 & 0.15 & 0.08 & 0.11 & 0.15 & 0.08 & 0.11 \\
100 & . & 4.94 & 4.96 & . & 97.40 & 97.96 & . & 0.07 & 0.07 & . & 0.07 & 0.07 \\
150 & . & . & 4.90 & . & . & 96.36 & . & . & 0.05 & . & . & 0.05 \\
  \hline \multicolumn{13}{l}{Number of Factors: 5} \\ \hline \hline \multicolumn{13}{l}{rigorous Lasso (HAC robust)} \\ \hline 
50 & 2.84 & 3.70 & 3.90 & 54.80 & 72.00 & 78.00 & 0.22 & 0.22 & 0.00 & 0.22 & 0.22 & 0.00 \\
100 & 3.02 & 3.60 & 3.98 & 60.40 & 72.00 & 79.20 & 0.00 & 0.00 & 0.02 & 0.00 & 0.00 & 0.02 \\
150 & 3.02 & 3.90 & 4.06 & 60.00 & 77.20 & 81.20 & 0.01 & 0.03 & 0.00 & 0.01 & 0.04 & 0.00 \\
\hline \multicolumn{13}{l}{Adaptive Lasso (AIC)} \\ \hline 
50 & 4.68 & 4.49 & 4.66 & 89.48 & 89.72 & 93.20 & 0.47 & 0.00 & 0.00 & 0.45 & 0.00 & 0.00 \\
100 & 3.62 & 4.71 & 4.78 & 70.20 & 86.52 & 95.56 & 0.12 & 0.41 & 0.00 & 0.14 & 0.40 & 0.00 \\
150 & 3.91 & 3.83 & 4.42 & 73.16 & 73.20 & 81.20 & 0.17 & 0.12 & 0.25 & 0.19 & 0.13 & 0.25 \\
\hline \multicolumn{13}{l}{Adaptive Lasso (AICC)} \\ \hline 
50 & 4.53 & 4.48 & 4.65 & 83.88 & 89.52 & 93.08 & 0.74 & 0.00 & 0.00 & 0.71 & 0.00 & 0.00 \\
100 & 3.59 & 4.69 & 4.78 & 69.68 & 84.16 & 95.56 & 0.11 & 0.50 & 0.00 & 0.13 & 0.49 & 0.00 \\
150 & 3.87 & 3.83 & 4.44 & 72.60 & 73.08 & 80.56 & 0.17 & 0.12 & 0.28 & 0.18 & 0.14 & 0.28 \\
\hline \multicolumn{13}{l}{Adaptive Lasso (BIC)} \\ \hline 
50 & 4.53 & 4.34 & 4.59 & 83.88 & 86.80 & 91.76 & 0.75 & 0.00 & 0.00 & 0.72 & 0.00 & 0.00 \\
100 & 3.48 & 4.77 & 4.71 & 67.76 & 81.04 & 94.28 & 0.09 & 0.76 & 0.00 & 0.11 & 0.74 & 0.00 \\
150 & 3.75 & 3.75 & 4.61 & 70.80 & 71.88 & 74.72 & 0.14 & 0.11 & 0.61 & 0.16 & 0.13 & 0.60 \\
\hline \multicolumn{13}{l}{Brownlees and Meesters} \\ \hline 
50 & 4.84 & 4.88 & 4.86 & 96.56 & 97.48 & 97.08 & 0.02 & 0.01 & 0.01 & 0.02 & 0.01 & 0.01 \\
100 & . & 4.87 & 4.85 & . & 96.84 & 96.56 & . & 0.03 & 0.02 & . & 0.02 & 0.02 \\
150 & . & . & 4.78 & . & . & 95.16 & . & . & 0.01 & . & . & 0.01 \\
\hline \hline \end{tabular} \caption{Monte Carlo Results for estimation of Dominant Drivers, Specification 4. The number of dominant Units is 5. See notes of Table \ref{tab:tab_speclasso1_5_5} and Table \ref{tab:tab_spec1_5_5}.} \label{tab:tab_spec4_5_5} \end{table}

\clearpage
 \begin{table}[h]     \scriptsize     \centering     \begin{tabular}{@{\extracolsep{4pt}} c c c c c c c c c c c c c c}     \hline \hline  & &\multicolumn{3}{c}{\(\hat{N_d}\)} & \multicolumn{3}{c}{TPR} & \multicolumn{3}{c}{FPR} & \multicolumn{3}{c}{FDR} \\ \cline{3-5} \cline{6-8} \cline{9-11} \cline{12-14} N/T & \(N_d\) & 50 &100 &150 & 50 &100 &150 & 50 &100 &150 & 50 &100 &150 \\ \hline \hline \multicolumn{14}{l}{Number of Factors: 0} \\ \hline \hline \multicolumn{14}{l}{rigorous LASSO (HAC robust)} \\ \hline 
50 & 5 & 3.28 & 4.58 & 4.93 & 63.80 & 91.40 & 98.18 & 0.20 & 0.02 & 0.04 & 0.21 & 0.03 & 0.04 \\
100 & 10 & 4.95 & 7.70 & 8.22 & 49.26 & 77.00 & 82.20 & 0.02 & 0.00 & 0.00 & 0.04 & 0.00 & 0.00 \\
150 & 15 & 7.48 & 10.05 & 11.66 & 49.80 & 66.93 & 77.73 & 0.01 & 0.01 & 0.00 & 0.01 & 0.01 & 0.00 \\
\hline  \multicolumn{14}{l}{Adaptive LASSO (AIC)} \\ \hline 
50 & 5 & 2.49 & 4.98 & 5.00 & 42.48 & 99.68 & 100.00 & 0.82 & 0.00 & 0.00 & 2.17 & 0.00 & 0.00 \\
100 & 10 & 9.58 & 4.06 & 10.00 & 95.60 & 39.18 & 100.00 & 0.02 & 0.16 & 0.00 & 0.02 & 0.23 & 0.00 \\
150 & 15 & 13.36 & 14.96 & 6.28 & 88.33 & 99.75 & 41.56 & 0.08 & 0.00 & 0.04 & 0.08 & 0.00 & 0.04 \\
\hline \multicolumn{14}{l}{Adaptive LASSO (AICC)} \\ \hline 
50 & 5 & 2.48 & 5.00 & 5.00 & 40.72 & 100.00 & 100.00 & 0.98 & 0.00 & 0.00 & 2.82 & 0.00 & 0.00 \\
100 & 10 & 9.81 & 4.01 & 10.00 & 97.88 & 39.10 & 100.00 & 0.03 & 0.11 & 0.00 & 0.03 & 0.21 & 0.00 \\
150 & 15 & 14.32 & 15.00 & 6.02 & 94.61 & 100.00 & 39.96 & 0.09 & 0.00 & 0.02 & 0.15 & 0.00 & 0.03 \\
\hline \multicolumn{14}{l}{Adaptive LASSO (BIC)} \\ \hline 
50 & 5 & 2.44 & 5.00 & 5.00 & 42.48 & 100.00 & 100.00 & 0.69 & 0.00 & 0.00 & 2.17 & 0.00 & 0.00 \\
100 & 10 & 9.70 & 4.31 & 10.00 & 96.66 & 42.02 & 100.00 & 0.03 & 0.12 & 0.00 & 0.03 & 0.17 & 0.00 \\
150 & 15 & 13.80 & 14.96 & 9.55 & 91.64 & 99.65 & 63.39 & 0.04 & 0.01 & 0.03 & 0.05 & 0.01 & 0.03 \\
\hline \multicolumn{14}{l}{BM procedure} \\ \hline 
50 & 5 & 4.99 & 5.00 & 5.00 & 99.60 & 100.00 & 100.00 & 0.02 & 0.00 & 0.00 & 0.02 & 0.00 & 0.00 \\
100 & 10 & . & 10.00 & 10.00 & . & 100.00 & 100.00 & . & 0.00 & 0.00 & . & 0.00 & 0.00 \\
150 & 15 & . & . & 15.00 & . & . & 100.00 & . & . & 0.00 & . & . & 0.00 \\
  \hline \multicolumn{14}{l}{Number of Factors: 1} \\ \hline \hline \multicolumn{14}{l}{rigorous LASSO (HAC robust)} \\ \hline 
50 & 5 & 2.89 & 4.13 & 4.46 & 56.60 & 82.20 & 89.20 & 0.13 & 0.04 & 0.00 & 0.19 & 0.04 & 0.00 \\
100 & 10 & 6.61 & 8.50 & 9.22 & 65.90 & 84.80 & 92.10 & 0.02 & 0.02 & 0.01 & 0.02 & 0.02 & 0.01 \\
150 & 15 & 10.54 & 13.36 & 13.55 & 70.27 & 88.87 & 90.27 & 0.00 & 0.02 & 0.01 & 0.00 & 0.02 & 0.01 \\
\hline  \multicolumn{14}{l}{Adaptive LASSO (AIC)} \\ \hline 
50 & 5 & 2.50 & 4.94 & 5.00 & 41.24 & 98.76 & 100.00 & 0.98 & 0.01 & 0.00 & 2.01 & 0.01 & 0.00 \\
100 & 10 & 9.77 & 7.58 & 10.00 & 97.66 & 75.32 & 99.98 & 0.00 & 0.05 & 0.00 & 0.00 & 0.05 & 0.00 \\
150 & 15 & 14.79 & 14.88 & 14.97 & 98.52 & 99.13 & 99.80 & 0.01 & 0.01 & 0.00 & 0.01 & 0.01 & 0.00 \\
\hline \multicolumn{14}{l}{Adaptive LASSO (AICC)} \\ \hline 
50 & 5 & 2.60 & 5.00 & 5.00 & 41.08 & 99.92 & 100.00 & 1.20 & 0.00 & 0.00 & 2.73 & 0.00 & 0.00 \\
100 & 10 & 9.76 & 9.54 & 10.00 & 97.34 & 95.26 & 100.00 & 0.03 & 0.01 & 0.00 & 0.03 & 0.01 & 0.00 \\
150 & 15 & 14.74 & 14.87 & 14.96 & 98.13 & 99.07 & 99.72 & 0.02 & 0.01 & 0.00 & 0.02 & 0.01 & 0.00 \\
\hline \multicolumn{14}{l}{Adaptive LASSO (BIC)} \\ \hline 
50 & 5 & 2.88 & 4.98 & 5.00 & 49.92 & 99.68 & 100.00 & 0.84 & 0.00 & 0.00 & 1.30 & 0.00 & 0.00 \\
100 & 10 & 9.72 & 9.89 & 10.00 & 96.88 & 98.80 & 100.00 & 0.03 & 0.01 & 0.00 & 0.03 & 0.01 & 0.00 \\
150 & 15 & 14.71 & 14.83 & 14.94 & 97.97 & 98.76 & 99.59 & 0.01 & 0.01 & 0.00 & 0.01 & 0.01 & 0.00 \\
\hline \multicolumn{14}{l}{BM procedure} \\ \hline 
50 & 5 & 4.46 & 4.77 & 4.75 & 88.92 & 95.16 & 94.84 & 0.04 & 0.02 & 0.02 & 0.03 & 0.02 & 0.02 \\
100 & 10 & . & 9.99 & 10.00 & . & 99.92 & 99.96 & . & 0.00 & 0.00 & . & 0.00 & 0.00 \\
150 & 15 & . & . & 15.00 & . & . & 99.99 & . & . & 0.00 & . & . & 0.00 \\
  \hline \multicolumn{14}{l}{Number of Factors: 5} \\ \hline \hline \multicolumn{14}{l}{rigorous LASSO (HAC robust)} \\ \hline 
50 & 5 & 4.88 & 4.97 & 5.00 & 97.00 & 99.40 & 100.00 & 0.07 & 0.00 & 0.00 & 0.07 & 0.00 & 0.00 \\
100 & 10 & 10.00 & 10.01 & 10.00 & 99.90 & 100.00 & 100.00 & 0.01 & 0.01 & 0.00 & 0.01 & 0.01 & 0.00 \\
150 & 15 & 15.00 & 15.00 & 14.98 & 100.00 & 35.07 & 33.99 & 0.00 & 6.06 & 5.84 & 0.00 & 38.78 & 33.93 \\
\hline  \multicolumn{14}{l}{Adaptive LASSO (AIC)} \\ \hline 
50 & 5 & 5.09 & 5.00 & 5.00 & 99.96 & 99.88 & 100.00 & 0.20 & 0.01 & 0.00 & 0.20 & 0.01 & 0.00 \\
100 & 10 & 6.64 & 9.56 & 10.00 & 66.36 & 95.62 & 100.00 & 0.00 & 0.00 & 0.00 & 0.00 & 0.00 & 0.00 \\
150 & 15 & 8.64 & 9.05 & 14.70 & 57.61 & 60.31 & 98.01 & 0.00 & 0.00 & 0.00 & 0.00 & 0.00 & 0.00 \\
\hline \multicolumn{14}{l}{Adaptive LASSO (AICC)} \\ \hline 
50 & 5 & 5.10 & 5.00 & 5.00 & 100.00 & 99.88 & 100.00 & 0.21 & 0.01 & 0.00 & 0.21 & 0.01 & 0.00 \\
100 & 10 & 6.56 & 9.58 & 10.00 & 65.58 & 95.78 & 100.00 & 0.00 & 0.00 & 0.00 & 0.00 & 0.00 & 0.00 \\
150 & 15 & 8.44 & 9.03 & 14.70 & 56.29 & 60.21 & 98.03 & 0.00 & 0.00 & 0.00 & 0.00 & 0.00 & 0.00 \\
\hline \multicolumn{14}{l}{Adaptive LASSO (BIC)} \\ \hline 
50 & 5 & 5.06 & 4.99 & 5.00 & 100.00 & 99.80 & 100.00 & 0.14 & 0.01 & 0.00 & 0.14 & 0.01 & 0.00 \\
100 & 10 & 6.43 & 9.56 & 10.00 & 64.28 & 95.58 & 100.00 & 0.00 & 0.00 & 0.00 & 0.00 & 0.00 & 0.00 \\
150 & 15 & 8.27 & 8.67 & 14.51 & 55.16 & 57.79 & 96.76 & 0.00 & 0.00 & 0.00 & 0.00 & 0.00 & 0.00 \\
\hline \multicolumn{14}{l}{BM procedure} \\ \hline 
50 & 5 & 4.81 & 4.84 & 4.83 & 95.92 & 96.60 & 96.24 & 0.02 & 0.02 & 0.04 & 0.02 & 0.02 & 0.04 \\
100 & 10 & . & 9.58 & 9.25 & . & 95.78 & 92.54 & . & 0.00 & 0.00 & . & 0.00 & 0.00 \\
150 & 15 & . & . & 13.93 & . & . & 92.89 & . & . & 0.00 & . & . & 0.00 \\
\hline \hline \end{tabular} \caption{Monte Carlo Results for estimation of Dominant Units: Number of dominant Units is \(\lceil0.1N\rceil\), Specification 5. See notes of Table \ref{tab:tab_speclasso1_5_5} and Table \ref{tab:tab_spec1_5_5}.} \label{tab:tab_spec5_1_5} \end{table}
 \begin{table}[h]     \scriptsize     \centering     \begin{tabular}{@{\extracolsep{4pt}} c c c c c c c c c c c c c c}     \hline \hline  & &\multicolumn{3}{c}{\(\hat{N_d}\)} & \multicolumn{3}{c}{TPR} & \multicolumn{3}{c}{FPR} & \multicolumn{3}{c}{FDR} \\ \cline{3-5} \cline{6-8} \cline{9-11} \cline{12-14} N/T & \(N_d\) & 50 &100 &150 & 50 &100 &150 & 50 &100 &150 & 50 &100 &150 \\ \hline \hline \multicolumn{14}{l}{Number of Factors: 0} \\ \hline \hline \multicolumn{14}{l}{rigorous LASSO (HAC robust)} \\ \hline 
50 & 25 & 7.26 & 10.73 & 14.35 & 28.52 & 42.48 & 57.28 & 0.52 & 0.44 & 0.12 & 0.71 & 0.61 & 0.16 \\
100 & 50 & 16.92 & 24.63 & 28.60 & 33.72 & 49.22 & 57.18 & 0.12 & 0.04 & 0.02 & 0.18 & 0.06 & 0.03 \\
150 & 75 & 26.88 & 36.89 & 39.14 & 35.84 & 49.17 & 52.17 & 0.00 & 0.01 & 0.01 & 0.00 & 0.02 & 0.02 \\
\hline  \multicolumn{14}{l}{Adaptive LASSO (AIC)} \\ \hline 
50 & 25 & 6.47 & 24.82 & 24.96 & 25.68 & 99.28 & 99.82 & 0.22 & 0.00 & 0.00 & 0.35 & 0.00 & 0.00 \\
100 & 50 & 40.11 & 10.39 & 49.97 & 80.21 & 20.73 & 99.94 & 0.00 & 0.04 & 0.00 & 0.00 & 0.06 & 0.00 \\
150 & 75 & 60.00 & 73.29 & 19.76 & 80.01 & 97.72 & 26.33 & 0.00 & 0.00 & 0.01 & 0.00 & 0.00 & 0.01 \\
\hline \multicolumn{14}{l}{Adaptive LASSO (AICC)} \\ \hline 
50 & 25 & 7.25 & 24.89 & 24.96 & 28.50 & 99.57 & 99.86 & 0.50 & 0.00 & 0.00 & 0.81 & 0.00 & 0.00 \\
100 & 50 & 12.29 & 15.93 & 50.00 & 24.53 & 31.77 & 100.00 & 0.04 & 0.10 & 0.00 & 0.05 & 0.16 & 0.00 \\
150 & 75 & 6.95 & 32.29 & 54.46 & 9.23 & 43.05 & 72.61 & 0.04 & 0.00 & 0.00 & 0.05 & 0.00 & 0.00 \\
\hline \multicolumn{14}{l}{Adaptive LASSO (BIC)} \\ \hline 
50 & 25 & 6.63 & 24.85 & 24.96 & 26.31 & 99.39 & 99.83 & 0.21 & 0.00 & 0.00 & 0.33 & 0.00 & 0.00 \\
100 & 50 & 21.30 & 13.56 & 49.99 & 42.59 & 27.06 & 99.98 & 0.01 & 0.05 & 0.00 & 0.01 & 0.07 & 0.00 \\
150 & 75 & 33.50 & 13.64 & 36.51 & 44.67 & 18.19 & 48.69 & 0.00 & 0.00 & 0.00 & 0.00 & 0.00 & 0.00 \\
\hline \multicolumn{14}{l}{BM procedure} \\ \hline 
50 & 25 & 23.12 & 24.93 & 24.98 & 92.49 & 99.71 & 99.93 & 0.00 & 0.00 & 0.00 & 0.00 & 0.00 & 0.00 \\
100 & 50 & . & 49.86 & 50.00 & . & 99.71 & 100.00 & . & 0.00 & 0.00 & . & 0.00 & 0.00 \\
150 & 75 & . & . & 75.00 & . & . & 100.00 & . & . & 0.00 & . & . & 0.00 \\
  \hline \multicolumn{14}{l}{Number of Factors: 1} \\ \hline \hline \multicolumn{14}{l}{rigorous LASSO (HAC robust)} \\ \hline 
50 & 25 & 8.64 & 12.85 & 16.98 & 34.32 & 51.36 & 67.92 & 0.24 & 0.04 & 0.00 & 0.34 & 0.05 & 0.00 \\
100 & 50 & 16.05 & 23.35 & 29.23 & 32.08 & 46.70 & 58.46 & 0.02 & 0.00 & 0.00 & 0.03 & 0.00 & 0.00 \\
150 & 75 & 19.07 & 28.28 & 33.41 & 25.43 & 37.71 & 44.55 & 0.00 & 0.00 & 0.00 & 0.00 & 0.00 & 0.00 \\
\hline  \multicolumn{14}{l}{Adaptive LASSO (AIC)} \\ \hline 
50 & 25 & 17.64 & 24.28 & 24.55 & 70.51 & 97.12 & 98.20 & 0.05 & 0.00 & 0.00 & 0.05 & 0.00 & 0.00 \\
100 & 50 & 28.05 & 49.79 & 49.89 & 56.10 & 99.58 & 99.78 & 0.00 & 0.00 & 0.00 & 0.00 & 0.00 & 0.00 \\
150 & 75 & 17.73 & 37.87 & 74.68 & 23.64 & 50.49 & 99.57 & 0.00 & 0.00 & 0.00 & 0.00 & 0.00 & 0.00 \\
\hline \multicolumn{14}{l}{Adaptive LASSO (AICC)} \\ \hline 
50 & 25 & 20.26 & 24.30 & 24.57 & 81.02 & 97.18 & 98.28 & 0.00 & 0.00 & 0.00 & 0.00 & 0.00 & 0.00 \\
100 & 50 & 15.78 & 46.82 & 49.99 & 31.56 & 93.63 & 99.98 & 0.00 & 0.00 & 0.00 & 0.00 & 0.00 & 0.00 \\
150 & 75 & 15.78 & 30.30 & 71.64 & 21.04 & 40.41 & 95.52 & 0.00 & 0.00 & 0.00 & 0.00 & 0.00 & 0.00 \\
\hline \multicolumn{14}{l}{Adaptive LASSO (BIC)} \\ \hline 
50 & 25 & 20.67 & 24.12 & 24.38 & 82.67 & 96.48 & 97.52 & 0.02 & 0.00 & 0.00 & 0.02 & 0.00 & 0.00 \\
100 & 50 & 23.95 & 48.77 & 49.99 & 47.90 & 97.54 & 99.98 & 0.00 & 0.00 & 0.00 & 0.00 & 0.00 & 0.00 \\
150 & 75 & 19.07 & 34.23 & 71.14 & 25.42 & 45.64 & 94.85 & 0.00 & 0.00 & 0.00 & 0.00 & 0.00 & 0.00 \\
\hline \multicolumn{14}{l}{BM procedure} \\ \hline 
50 & 25 & 8.84 & 10.15 & 11.39 & 35.36 & 40.58 & 45.56 & 0.00 & 0.00 & 0.00 & 0.00 & 0.00 & 0.00 \\
100 & 50 & . & 13.81 & 16.52 & . & 27.62 & 33.04 & . & 0.00 & 0.00 & . & 0.00 & 0.00 \\
150 & 75 & . & . & 19.53 & . & . & 26.04 & . & . & 0.00 & . & . & 0.00 \\
  \hline \multicolumn{14}{l}{Number of Factors: 5} \\ \hline \hline \multicolumn{14}{l}{rigorous LASSO (HAC robust)} \\ \hline 
50 & 25 & 24.94 & 25.00 & 25.00 & 69.98 & 67.87 & 66.95 & 41.21 & 41.17 & 41.13 & 96.90 & 83.80 & 78.89 \\
100 & 50 & 24.94 & 50.00 & 50.00 & 34.19 & 69.17 & 69.99 & 21.55 & 40.73 & 40.05 & 59.90 & 84.81 & 80.92 \\
150 & 75 & 22.46 & 74.98 & 75.00 & 16.97 & 64.89 & 67.07 & 13.87 & 43.05 & 41.64 & 25.06 & 93.04 & 85.45 \\
\hline  \multicolumn{14}{l}{Adaptive LASSO (AIC)} \\ \hline 
50 & 25 & 13.94 & 18.41 & 20.72 & 55.77 & 73.63 & 82.89 & 0.00 & 0.00 & 0.00 & 0.00 & 0.00 & 0.00 \\
100 & 50 & 11.81 & 22.00 & 23.69 & 23.62 & 44.00 & 47.37 & 0.00 & 0.00 & 0.00 & 0.00 & 0.00 & 0.00 \\
150 & 75 & 13.41 & 14.61 & 28.11 & 17.88 & 19.48 & 37.49 & 0.00 & 0.00 & 0.00 & 0.00 & 0.00 & 0.00 \\
\hline \multicolumn{14}{l}{Adaptive LASSO (AICC)} \\ \hline 
50 & 25 & 12.84 & 17.87 & 20.53 & 51.38 & 71.50 & 82.10 & 0.00 & 0.00 & 0.00 & 0.00 & 0.00 & 0.00 \\
100 & 50 & 11.32 & 22.05 & 22.93 & 22.63 & 44.11 & 45.87 & 0.00 & 0.00 & 0.00 & 0.00 & 0.00 & 0.00 \\
150 & 75 & 12.64 & 14.58 & 27.78 & 16.85 & 19.44 & 37.04 & 0.00 & 0.00 & 0.00 & 0.00 & 0.00 & 0.00 \\
\hline \multicolumn{14}{l}{Adaptive LASSO (BIC)} \\ \hline 
50 & 25 & 13.05 & 16.22 & 19.20 & 52.19 & 64.87 & 76.80 & 0.00 & 0.00 & 0.00 & 0.00 & 0.00 & 0.00 \\
100 & 50 & 11.25 & 21.12 & 19.69 & 22.51 & 42.24 & 39.39 & 0.00 & 0.00 & 0.00 & 0.00 & 0.00 & 0.00 \\
150 & 75 & 12.83 & 14.28 & 27.72 & 17.10 & 19.05 & 36.96 & 0.00 & 0.00 & 0.00 & 0.00 & 0.00 & 0.00 \\
\hline \multicolumn{14}{l}{BM procedure} \\ \hline 
50 & 25 & 3.64 & 3.58 & 3.74 & 14.55 & 14.34 & 14.97 & 0.00 & 0.00 & 0.00 & 0.00 & 0.00 & 0.00 \\
100 & 50 & . & 2.64 & 3.07 & . & 5.28 & 6.13 & . & 0.00 & 0.00 & . & 0.00 & 0.00 \\
150 & 75 & . & . & 2.83 & . & . & 3.78 & . & . & 0.00 & . & . & 0.00 \\
\hline \hline \end{tabular} \caption{Monte Carlo Results for estimation of Dominant Units: Number of dominant Units is \(\lceil0.5N\rceil\), Specification 5. See notes of Table \ref{tab:tab_speclasso1_5_5} and Table \ref{tab:tab_spec1_5_5}.} \label{tab:tab_spec5_2_5} \end{table}
 \begin{table}[h]     \scriptsize     \centering     \begin{tabular}{@{\extracolsep{4pt}} c c c c c c c c c c c c c c}     \hline \hline  & &\multicolumn{3}{c}{\(\hat{N_d}\)} & \multicolumn{3}{c}{TPR} & \multicolumn{3}{c}{FPR} & \multicolumn{3}{c}{FDR} \\ \cline{3-5} \cline{6-8} \cline{9-11} \cline{12-14} N/T & \(N_d\) & 50 &100 &150 & 50 &100 &150 & 50 &100 &150 & 50 &100 &150 \\ \hline \hline \multicolumn{14}{l}{Number of Factors: 0} \\ \hline \hline \multicolumn{14}{l}{rigorous LASSO (HAC robust)} \\ \hline 
50 & 45 & 6.35 & 7.69 & 7.66 & 13.53 & 16.16 & 16.33 & 5.20 & 8.40 & 6.20 & 5.38 & 8.11 & 5.26 \\
100 & 90 & 8.94 & 9.71 & 11.67 & 9.91 & 10.73 & 12.88 & 0.20 & 0.50 & 0.80 & 0.36 & 0.81 & 1.34 \\
150 & 135 & 8.50 & 8.78 & 14.50 & 6.29 & 6.50 & 10.74 & 0.07 & 0.00 & 0.00 & 0.14 & 0.00 & 0.00 \\
\hline  \multicolumn{14}{l}{Adaptive LASSO (AIC)} \\ \hline 
50 & 45 & 5.12 & 5.13 & 6.07 & 11.38 & 11.40 & 13.48 & 0.08 & 0.00 & 0.00 & 0.12 & 0.00 & 0.00 \\
100 & 90 & 3.72 & 3.94 & 4.49 & 4.13 & 4.38 & 4.99 & 0.00 & 0.00 & 0.00 & 0.00 & 0.00 & 0.00 \\
150 & 135 & 4.00 & 3.31 & 3.08 & 2.96 & 2.45 & 2.28 & 0.00 & 0.00 & 0.00 & 0.00 & 0.00 & 0.00 \\
\hline \multicolumn{14}{l}{Adaptive LASSO (AICC)} \\ \hline 
50 & 45 & 5.35 & 5.23 & 6.23 & 11.85 & 11.62 & 13.84 & 0.32 & 0.00 & 0.00 & 0.46 & 0.00 & 0.00 \\
100 & 90 & 3.97 & 3.99 & 3.86 & 4.40 & 4.44 & 4.29 & 0.06 & 0.00 & 0.00 & 0.15 & 0.00 & 0.00 \\
150 & 135 & 3.26 & 3.36 & 4.02 & 2.41 & 2.49 & 2.98 & 0.00 & 0.00 & 0.00 & 0.00 & 0.00 & 0.00 \\
\hline \multicolumn{14}{l}{Adaptive LASSO (BIC)} \\ \hline 
50 & 45 & 4.78 & 6.16 & 6.36 & 10.60 & 13.68 & 14.13 & 0.20 & 0.00 & 0.00 & 0.29 & 0.00 & 0.00 \\
100 & 90 & 3.84 & 4.39 & 4.08 & 4.27 & 4.88 & 4.54 & 0.00 & 0.00 & 0.00 & 0.00 & 0.00 & 0.00 \\
150 & 135 & 3.75 & 4.75 & 2.99 & 2.78 & 3.51 & 2.21 & 0.00 & 0.09 & 0.00 & 0.00 & 0.18 & 0.00 \\
\hline \multicolumn{14}{l}{BM procedure} \\ \hline 
50 & 45 & 3.29 & 3.65 & 4.49 & 7.32 & 8.11 & 9.97 & 0.00 & 0.00 & 0.00 & 0.00 & 0.00 & 0.00 \\
100 & 90 & . & 3.01 & 3.12 & . & 3.34 & 3.47 & . & 0.00 & 0.00 & . & 0.00 & 0.00 \\
150 & 135 & . & . & 2.74 & . & . & 2.03 & . & . & 0.00 & . & . & 0.00 \\
  \hline \multicolumn{14}{l}{Number of Factors: 1} \\ \hline \hline \multicolumn{14}{l}{rigorous LASSO (HAC robust)} \\ \hline 
50 & 45 & 5.50 & 6.91 & 5.29 & 12.16 & 15.33 & 11.73 & 0.60 & 0.20 & 0.20 & 0.85 & 0.27 & 0.28 \\
100 & 90 & 6.04 & 5.44 & 3.96 & 6.71 & 6.04 & 4.40 & 0.00 & 0.00 & 0.00 & 0.00 & 0.00 & 0.00 \\
150 & 135 & 4.55 & 3.22 & 4.22 & 3.37 & 2.39 & 3.13 & 0.00 & 0.00 & 0.00 & 0.00 & 0.00 & 0.00 \\
\hline  \multicolumn{14}{l}{Adaptive LASSO (AIC)} \\ \hline 
50 & 45 & 4.25 & 5.53 & 6.25 & 9.45 & 12.28 & 13.88 & 0.00 & 0.00 & 0.00 & 0.00 & 0.00 & 0.00 \\
100 & 90 & 4.35 & 3.68 & 4.81 & 4.83 & 4.08 & 5.34 & 0.00 & 0.00 & 0.00 & 0.00 & 0.00 & 0.00 \\
150 & 135 & 5.27 & 4.46 & 3.21 & 3.91 & 3.31 & 2.38 & 0.00 & 0.00 & 0.00 & 0.00 & 0.00 & 0.00 \\
\hline \multicolumn{14}{l}{Adaptive LASSO (AICC)} \\ \hline 
50 & 45 & 5.84 & 5.79 & 5.81 & 12.97 & 12.86 & 12.92 & 0.00 & 0.00 & 0.00 & 0.00 & 0.00 & 0.00 \\
100 & 90 & 8.78 & 6.29 & 4.46 & 9.75 & 6.98 & 4.96 & 0.00 & 0.00 & 0.00 & 0.00 & 0.00 & 0.00 \\
150 & 135 & 11.50 & 10.30 & 8.10 & 8.52 & 7.63 & 6.00 & 0.00 & 0.00 & 0.00 & 0.00 & 0.00 & 0.00 \\
\hline \multicolumn{14}{l}{Adaptive LASSO (BIC)} \\ \hline 
50 & 45 & 4.49 & 3.95 & 3.89 & 9.97 & 8.78 & 8.64 & 0.00 & 0.00 & 0.00 & 0.00 & 0.00 & 0.00 \\
100 & 90 & 5.57 & 6.30 & 3.13 & 6.19 & 7.00 & 3.47 & 0.00 & 0.00 & 0.00 & 0.00 & 0.00 & 0.00 \\
150 & 135 & 5.48 & 8.47 & 4.96 & 4.06 & 6.28 & 3.68 & 0.00 & 0.00 & 0.00 & 0.00 & 0.00 & 0.00 \\
\hline \multicolumn{14}{l}{BM procedure} \\ \hline 
50 & 45 & 5.11 & 8.11 & 9.69 & 11.36 & 18.01 & 21.52 & 0.00 & 0.00 & 0.00 & 0.00 & 0.00 & 0.00 \\
100 & 90 & . & 9.52 & 10.15 & . & 10.58 & 11.27 & . & 0.00 & 0.00 & . & 0.00 & 0.00 \\
150 & 135 & . & . & 11.23 & . & . & 8.32 & . & . & 0.00 & . & . & 0.00 \\
  \hline \multicolumn{14}{l}{Number of Factors: 5} \\ \hline \hline \multicolumn{14}{l}{rigorous LASSO (HAC robust)} \\ \hline 
50 & 45 & 34.42 & 43.20 & 44.98 & 74.51 & 92.76 & 95.81 & 67.07 & 83.81 & 87.41 & 216.96 & 207.49 & 201.28 \\
100 & 90 & 19.36 & 86.44 & 88.18 & 20.96 & 92.65 & 94.87 & 19.01 & 84.80 & 86.13 & 89.32 & 322.03 & 281.95 \\
150 & 135 & 4.70 & 39.76 & 126.94 & 3.72 & 28.15 & 90.03 & 3.02 & 26.15 & 83.30 & 16.86 & 119.87 & 340.04 \\
\hline  \multicolumn{14}{l}{Adaptive LASSO (AIC)} \\ \hline 
50 & 45 & 7.88 & 5.64 & 5.85 & 17.50 & 12.54 & 13.00 & 0.00 & 0.00 & 0.00 & 0.00 & 0.00 & 0.00 \\
100 & 90 & 13.33 & 16.44 & 5.18 & 14.81 & 18.26 & 5.76 & 0.00 & 0.00 & 0.00 & 0.00 & 0.00 & 0.00 \\
150 & 135 & 15.07 & 16.89 & 22.62 & 11.16 & 12.51 & 16.76 & 0.00 & 0.00 & 0.00 & 0.00 & 0.00 & 0.00 \\
\hline \multicolumn{14}{l}{Adaptive LASSO (AICC)} \\ \hline 
50 & 45 & 11.67 & 5.92 & 6.42 & 25.92 & 13.16 & 14.27 & 0.00 & 0.00 & 0.00 & 0.00 & 0.00 & 0.00 \\
100 & 90 & 12.45 & 20.06 & 6.13 & 13.83 & 22.28 & 6.81 & 0.00 & 0.00 & 0.00 & 0.00 & 0.00 & 0.00 \\
150 & 135 & 14.17 & 16.81 & 25.56 & 10.49 & 12.45 & 18.93 & 0.00 & 0.00 & 0.00 & 0.00 & 0.00 & 0.00 \\
\hline \multicolumn{14}{l}{Adaptive LASSO (BIC)} \\ \hline 
50 & 45 & 11.40 & 5.99 & 5.65 & 25.32 & 13.31 & 12.56 & 0.00 & 0.00 & 0.00 & 0.00 & 0.00 & 0.00 \\
100 & 90 & 12.67 & 19.86 & 5.36 & 14.08 & 22.06 & 5.95 & 0.00 & 0.00 & 0.00 & 0.00 & 0.00 & 0.00 \\
150 & 135 & 14.50 & 16.58 & 26.51 & 10.74 & 12.28 & 19.64 & 0.00 & 0.00 & 0.00 & 0.00 & 0.00 & 0.00 \\
\hline \multicolumn{14}{l}{BM procedure} \\ \hline 
50 & 45 & 2.58 & 2.67 & 2.61 & 5.72 & 5.94 & 5.79 & 0.00 & 0.00 & 0.00 & 0.00 & 0.00 & 0.00 \\
100 & 90 & . & 2.19 & 2.32 & . & 2.43 & 2.58 & . & 0.00 & 0.00 & . & 0.00 & 0.00 \\
150 & 135 & . & . & 2.17 & . & . & 1.61 & . & . & 0.00 & . & . & 0.00 \\
\hline \hline \end{tabular} \caption{Monte Carlo Results for estimation of Dominant Units: Number of dominant Units is \(\lceil0.9N\rceil\), Specification 5. See notes of Table \ref{tab:tab_speclasso1_5_5} and Table \ref{tab:tab_spec1_5_5}.} \label{tab:tab_spec5_3_5} \end{table}

\clearpage
\section{Empirical Application}

\begin{table}[!h]
\begin{tabular}{l l l l}\hline\hline
Abbreviation & Countryname & \\ \hline
arg&	Argentina &mex&	mexico\\
austlia&	Australia&neth&	Netherlands\\
austria&	Austria&nor&	Norway\\
bel	&Belgium&nzld&	New zealand\\
bra	&Brazil&per&	Peru\\
can	&Canada&phlp&	Philippines\\
china&China&safrc&	South Africa\\
chl	&Chile&sarbia&	Saudi Arabia\\
fin	&Finland&sing&	Singapore\\
france&	France&spain&	Spain\\
germ&Germany&swe&	Sweden\\
india&	India&switz&	Switzerland\\
indns&	Indonesia&thai&	Thailand\\
italy&	Italy&turk&	Turkey\\
japan&	Japan&uk&	United Kingdom\\
kor&	Korea&usa&	United States \\
mal&	Malaysia& & \\ \hline\hline
\end{tabular}
\caption{Countries in the GVAR dataset.}
\label{lab:ctry}
\end{table}

\end{document}